\documentclass{article}

\usepackage{arxiv}

\usepackage[utf8]{inputenc} 
\usepackage[T1]{fontenc}    
\usepackage{hyperref}       
\usepackage{url}            
\usepackage{booktabs}       
\usepackage{amsfonts}       
\usepackage{amsmath}
\usepackage{amsthm}
\usepackage{nicefrac}       
\usepackage{microtype}      
\usepackage{lipsum}
\usepackage{graphicx}
\graphicspath{ {./images/} }

\usepackage{caption}
\usepackage{subcaption}
\usepackage{algorithm}
\usepackage{algorithmic}
\usepackage{longtable} 
\usepackage{threeparttable}

\newtheorem{theorem}{Theorem}[section]

\newtheorem*{theorem*}{Theorem}

\theoremstyle{definition}
\newtheorem{definition}[theorem]{Definition}

\title{Follow the Leader: Enhancing Systematic Trend-Following Using Network Momentum}

\author{
 Linze Li \\
  Imperial College London, UK \\
  \texttt{linze.li20@imperial.ac.uk} \\
   \And
 William Ferreira \\
 \\
  \texttt{william.ferreira.14@alumni.ucl.ac.uk} \\ \\
}

\begin{document}
\maketitle
\begin{abstract}
We present a systematic, trend-following strategy, applied to commodity futures markets, that combines univariate trend indicators with cross-sectional trend indicators that capture so-called {\em momentum spillover}, which can occur when there is a lead-lag relationship between the trending behaviour of different markets. Our strategy utilises two methods for detecting lead-lag relationships, with a method for computing {\em network momentum}, to produce a novel trend-following indicator. We use our new trend indicator to construct a portfolio whose performance we compare to a baseline model which uses only univariate indicators, and demonstrate statistically significant improvements in Sharpe ratio, skewness of returns, and downside performance, using synthetic bootstrapped data samples taken from time-series of actual prices.
\end{abstract}


\section{Introduction}
Trend-following is a well-known and popular strategy in finance \cite{hurst2017century}. The basic premise of trend-following is that price direction exhibits persistence: assets whose prices have been rising tend to continue to rise (and conversely assets with falling prices tend to continue to fall); therefore, an investor may realise profits by buying (resp. selling) assets whose price has recently risen (resp. fallen) in the expectation that the price trend will continue. One of the characteristic and desirable features of trend-following strategies is the positive skew in the distribution of returns: regular small losses are compensated by fewer, but much larger gains.

The persistence of market returns has been extensively studied from macroeconomic perspectives. For instance, industrial growth rates have a significant impact on momentum profits \cite{Liu2008momentumprofits}; investor behaviours, such as delayed information reception and asynchronous response timings, support the slow information diffusion hypothesis \cite{Hong1999aunifiedtheoryofunderraction, hou2007industry}. Behavioural biases like conservatism may also encourage premature selling or prolonged holding of assets \cite{barberis1998model}. Such persistence in market returns continues until significant deviations from price fundamentals eventually trigger a market reversal \cite{vayanos2013institutional}. The profitability of time series momentum strategies is demonstrated across various markets, showing that purchasing stocks that perform well in recent months and selling those that show poor returns results in higher profits \cite{jegadeesh1993returns, rouwenhorst1998international, chui2000momentum}. This profitability has been rigorously validated through statistical experiments to confirm it is not due to random chance 
\cite{jegadeesh2001profitability, hurst2017century}. 

Price momentum extends beyond individual markets. \cite{gebhardt2005stock} observes that high equity returns in one year can predict high corporate bond returns the following year despite bonds lacking inherent momentum. This `cross-sectional momentum spillover' is primarily due to the bond market's delayed reaction to equity performance, known as the `lead-lag effect'. Previous literature has identified multiple drivers for the lead-lag effect, including factors such as company size \cite{lo1990whenarecontrarianprofits}, institutional ownership levels \cite{badrinath1995ofshepherds}, analyst coverage \cite{brennan1993investment}, and industry affiliation \cite{moskowitz1999industries, grobys2018risk}. Numerous studies have explored systematic approaches to capturing the lead-lag effect. For instance, \cite{campbell1998econometrics} utilised the difference in the cross-correlation function based on Pearson correlation, while \cite{wu2010detecting} employed the signed normalised area under the curve of the cross-correlation function as an indicator. Further, \cite{bennett2022lead-lag} experimented with alternatives to Pearson correlation, such as Kendall rank correlation \cite{kendall1938new}, distance correlation \cite{szekely2007measuring}, and mutual information from discretised time series values \cite{fiedor2014information}. For additional methods for detecting lead-lag effects, see \cite{billio2012econometric, wang2017lead, marti2017exploring} 

After computing lead-lag metrics pairwise, many studies suggest employing ranking algorithms to identify assets most likely to lead or follow \cite{cartea2023detecting, bennett2022lead-lag, ali2020shared}. Positions for the followers are established based on the average of the leaders' performance. For example, if leaders exhibit a negative average return, followers are shorted in anticipation of a similar downward trend. To rank markets based on their lead-lag behaviour cross-sectionally, the literature employs methods such as RowSum Rank \cite{cartea2023detecting, zhang2023dtw, bennett2022lead-lag}, PageRank \cite{wu2010detecting, basnarkov2020lead}, and machine learning approaches like the Learning-to-rank algorithm \cite{tan2023spatio, poh2020building}. Another strategy involves clustering lead-lag metrics using algorithms such as clustering by industry sectors \cite{hou2007industry, billio2012econometric}, k-means \cite{zhang2023dtw}, and spectral and Hermitian clustering \cite{bennett2022lead-lag, cartea2023detecting}. This data-driven methodology, prevalent in existing studies \cite{basnarkov2020lead, wu2010detecting}, suggests establishing positions within lagging clusters based on the average performance of markets in the leading cluster. This approach is used to capitalise on trend-following opportunities or to construct opposite positions to counteract mean-reverting behaviours \cite{cartea2023detecting}.

To the best of our knowledge, quantitative research that measures the influence of leading markets on the performance of lagging markets is currently limited, especially when portfolios span multiple industries. It is pointed out in \cite{pu2023networkmomentumassetclasses} that, although previous studies have identified momentum spillover across various sectors—such as equities and bonds \cite{gebhardt2005stock, haesen2017momentum}, equities and foreign currencies \cite{declerck2019trend}, currency news and bonds in emerging markets \cite{yamani2021currency}, and between crude oil indexes and equities \cite{fernandez2023cross}—the absence of firm-like economic and fundamental linkages in commodities markets complicates the identification of connections, such as those between orange juice and natural gas.
Moreover, while existing studies predominantly focus on statistically examining the momentum spillover effect and use it as a market selection mechanism for trading followers in exchanges, they fail to quantify and aggregate this momentum spillover into a new trading signal for portfolio construction. To bridge this gap, the existing literature suggests leveraging network theory. For instance, \cite{yamamoto2021momentum} uses edge centrality to quantify the importance of supplier-customer relationships, and \cite{pu2023networkmomentumassetclasses} explores the `network momentum' spillover across industries, aggregating momentum from commodities, equities, bonds, and currencies to create a novel trading signal. Specifically, the latter research employs ideas from graph learning, treating each market as a node within a graph. This approach uses time series momentum features, including moving average crossover signals and exponentially weighted returns, as signal processes for each node. They then solve a convex optimisation problem to approximate the weighted graph adjacency matrix. The edges of this graph elucidate the complex relationships across markets, with the magnitude of each edge reflecting the strength of similarity in momentum features between market pairs. Following this model, the time series momentum of other markets is weighted and used in a linear regression to predict the next day's returns. Subsequently, a portfolio is constructed by assigning binary positions: $+1$ for a long position or $-1$ for a short position, based on the sign of the predicted future returns \cite{zhang2023dtw, tan2023spatio, poh2020building}. However, such binary betting on positions based on momentum direction may not be optimal because this approach could lead to a discontinuous model, losing both convexity and positive skewness in returns \cite{martin2023design}. Similarly, \cite{pu2023networkmomentumassetclasses} notes that their strategy may not adequately address risk characteristics, potentially increasing exposure to significant downside movements.

Our work makes three contributions: first, it combines multiple lead-lag detection mechanisms as an ensemble model for price trend detection. Second, it uses the outputs of the ensemble model as a low-dimension feature space to learn a graph-adjacency matrix, to generate a network momentum indicator. Third, our network momentum indicator is used to construct a realistic portfolio for trend-following commodity futures that is statistically significantly superior to a baseline model that utilises only univariate trend indicators, when compared using bootstrapped price trajectories sampled from real historical price data.

The remainder of the paper is organised as follows. Chapter \ref{lead_lag_matrix} presents the frameworks used to identify the lead-lag relationship. Chapter \ref{network_momentum_matrix} introduces our approach to constructing the network momentum matrix. Chapter \ref{methodology} explains the strategy setup. Chapter \ref{portfolio construction} explains the construction of our portfolio based on the momentum signals, and the performance analysis is reported in Chapter \ref{analysis}. We conclude the paper with final thoughts and future directions in Chapter \ref{conclusion}, and provide auxiliary data in the Appendix.

\section{Lead-lag matrix construction} \label{lead_lag_matrix}
We employ two recently proposed methods
to identify and quantify the lead-lag relationship between pairs of assets. The first method, based on the L\'evy area of pairwise market returns \cite{cartea2023detecting}, identifies both linear and nonlinear relationships at a fixed lag length, for example a one day lag. The second model utilises the dynamic time warping (DTW) algorithm \cite{vintsyuk1968speech} on pairwise market returns, as presented in \cite{zhang2023dtw, Varfis2001}. The DTW model relaxes the fixed lag assumption and adeptly handles non-synchronised time series of varying lengths. Building upon these foundations, we explore various advanced DTW algorithms to capture the co-movement between two markets' returns. We then calculate pairwise lead-lag scores for all combinations of market pairs and use them to construct the lead-lag matrix, which
contains values indicating the direction of each market's lead against another. The lead-lag matrix is skew-symmetric since its transpose equals its negative -- this can be seen from if market $m$ leads market $n$ by a lag of $\ell$ where $\ell$ has the same unit as the time, then market $n$ leads market $m$ by a lag of $-\ell$.

\subsection{Signature based L\'evy area}\label{levy}
The first lead-lag model applied is based on the L\'evy area enclosed by two time-series, which can be used to identify both linear and non-linear lead-lag relationships between them. To understand the concept of the L\'evy area, it is necessary to introduce the concept of {\em path}; for a full discussion of paths the reader is referred to \cite{gyurkó2014extractinginformationsignaturefinancial, chevyrev2016primersignaturemethodmachine}. A {\em path} is a continuous function $X$ on an interval $[a,b]$, such that $X: [a,b] \rightarrow \mathbb{R}^n$. Assuming that $[a, b]$ represents an interval of time, we write $X_t$ for $X(t)$, the value of $X$ at time $t\in[a,b]$; when $n=2$ we write $X_t = \{X^{1}_t, X^{2}_t\}$. We can view the time-series of returns of two assets as following a two-dimensional path, where $(X^{1}_t, X^{2}_t)$ represents the returns at time $t$ of asset 1 and asset 2 respectively. Let $S(X)^{i,j}_{a,t}$\footnote{$S(X)^{i,j}$ is the 2nd-level {\em signature} of the path} be defined as:
$$
S(X)^{i,j}_{a,t} = \int_{a<r<s<t}dX^{i}_r dX^{j}_s
$$
which can be used to define the L\'evy area, $A^{\text{L\'evy}}$, enclosed by path $\{X^1_t, X^2_t\}$:
$$
A_{1,2}^{\text{L\'evy}} = \frac{1}{2}\left(S(X)^{1,2}_{a,b} - S(X)^{2,1}_{a,b}\right)
$$
If an increase (resp. decrease) in the $X^1$ is followed by an increase (resp. decrease) in the $X^2$, then the L\'evy area enclosed by the two series and the chord connecting the two ends, is positive. Conversely, if the movements of $X^1$ and $X^2$ are in opposite directions, then the L\'evy area is negative. For discrete processes $X_t^1$ and $X_t^2$, the L\'evy area can be expressed as:
\begin{equation}\label{formula:levy_area_discrete}
    A_{1,2}^{\text{L\'evy}} := \frac{1}{2} \left(\sum_{a<s<b} (-X_s^i X_{s-1}^j + X_s^j X_{s-1}^i) + X_a^i (X_a^j - X_b^j) + X_a^j (X_b^i - X_a^i)\right).
\end{equation}

Let $\tilde{r}_{n,t}$ denote the {\em standardised} return of asset $n$ at time $t$, i.e. the return on the asset normalised to have zero mean and unit variance, and suppose the functional form of the relationship between the standardised returns on asset $m$ and asset $n$ is given by:
\begin{equation}\label{formula:levy_area_cartea}
    \tilde{r}_{n,t} = \beta_\ell f(\tilde{r}_{m,t-\ell}) + \epsilon_t,
\end{equation}
where $t-\ell$ represents $\ell$ time-steps prior to $t$, and $f$ is any continuous function, then the following result \cite[Theorem 1, page 9]{cartea2023detecting} links (\ref{formula:levy_area_cartea}) to the L\'evy area:

\begin{theorem*}\label{thm:levy_area_lead_lag}
~\cite[Theorem 1, page 9]{cartea2023detecting} Assume $\{X^i_\tau\}_{\tau=s}^t$ and $\{X^j_\tau\}_{\tau=s}^t$ are two independent random processes with zero mean, unit variance, and symmetric distribution, and both satisfy \eqref{formula:levy_area_cartea} over a time interval [s,t]. Then, the sign of the L\'evy area $A_{i,j}^{\text{L\'evy}}$ between $X^i_\tau$ and $X^j_\tau$ is the same as the sign of $\ell$ if and only if $\ell = \pm 1$. In addition, if $\ell = \pm 1$ and the third derivative of the function $f$ exists, there is a constant $K$ such that for all pairs ($i$,$j$), we have
\begin{equation}\label{formula:levy_area_theorem}
    E[A_{i,j}^{\text{L\'evy}} - K\beta_\ell] = \frac{M}{6}\beta_\ell E[\sum_{s<a<t}f^{'''}(\xi_{a-1}^j)(X_{a-1}^j)^4]
\end{equation}
for some constant $M$ and $|\xi_{a-1}^j| < |X_{a-1}^j|$.
\end{theorem*}
Thus, we may use the L\'evy area to detect lead-lag relationships in asset returns.

\subsection{Dynamic Time Warping}\label{dtw}
The second lead-lag model applied is based on dynamic time warping. Compared to the signature-based L\'evy area model proposed in Section \ref{levy}, this model differs in two major respects: firstly, it does not assume a prefixed lag $\ell$ between two market return series; instead, the alignment between the two series is dynamically determined by the algorithm. Secondly, this model can handle non-synchronised paired series, thus making the analysis of markets with different lengths of return series feasible. We contend that the variation in matched indices from the two paired series 
can identify the leader and the follower within these pairs.

The classical dynamic time warping (DTW) model effectively identifies the lead-lag relationship as shown in \cite{zhang2023dtw, Varfis2001}. Suppose $X$ and $Y$ are two time series with lengths $m$ and $n$, respectively:
$$
X = X_1, X_2, \ldots, X_m, \quad Y = Y_1, Y_2, \ldots, Y_n,
$$
DTW creates a {\em warping path}, $\mathcal{W}\{X, Y\}$ defined by:
\begin{equation}\label{formula:warping_path}
    \mathcal{W}\{X, Y\} = \{w_1, \cdots, w_n, \cdots, w_L\} \quad \operatorname{max}(m,n) \leq L \leq m + n -1.
\end{equation}
where each element $w_k=(i, j)_k$ indicating that, at the $k^\text{th}$ step in the warping path, $X_i$ should be aligned with $Y_j$. This warping path is constructed with three essential constraints:
\begin{itemize}
    \item Boundary conditions: $w_1=(1,1)$ and $w_L=(m, n)$. This ensures that the first and last elements in $X$ and $Y$ are matched respectively. This boundary condition can, however, be partly relaxed. We refer interested readers to \cite{stubinger2022multidtw} for details.
    \item Monotonicity condition: Given $w_l = (i_l, j_l)$ for $l\in [1,\cdots, L]$, $i_1 \leq i_2 \leq \cdots \leq i_L$ and $j_1 \leq j_2 \leq \cdots \leq j_L$. This condition enforces the mapping in chronological order. 
    \item Step size condition: $w_{l+1} - w_l \in \{(1,0), (0,1), (1,1)\} \text{ for } l \in [1, \ldots, L-1]$. This is sometimes called the continuity constraint. It allows only adjacent cell transitions within the warping path.
\end{itemize}
The cost of a warping path $\mathcal{W}\{X, Y\}$ is defined as 
$$
c_\mathcal{W}(X, Y) := \sum_{l=1}^L c_\text{local}(X_{i_l}, Y_{j_l}),
$$
where $c_\text{local}$ is a local cost measure measuring the dissimilarity between two points. Typically, this is chosen as the Euclidean distance \cite{zhang2023dtw, Varfis2001}. 

Among the various warping paths, the optimal path $\mathcal{W}^*$, which is also the best alignment, follows a path of minimal cost through $c_\mathcal{W}(X, Y)$. The DTW distance between $X$ and $Y$ is then defined as
$$
\begin{aligned}
    \operatorname{DTW(X, Y)} 
    &:= \operatorname{min}\{c_\mathcal{W}(X, Y) | \mathcal{W}~\text{is a warping path satisfying necessary constraints}\}.
\end{aligned}
$$
The optimal warping path can be computed by a dynamic programming algorithm \cite{muller2007dynamictimewarping}.

In \cite{Keogh2001ddtw} it mentions that DTW aligns two series solely on their coordinate values. It faces difficulties when the two time sequences have similar local shapes but differ in their values -- a single point on one time series is mapped to several points on the other time series. This undesirable behaviour is referred to as `singularities'. According to \cite[page 2]{Keogh2001ddtw}, this occurs because the algorithm may attempt to account for variability in the Y-axis by warping the X-axis. Although one could, and in fact should, as stated by \cite{rakthanmanon2013addressing}, always perform Z-normalisation to convert the time sequences to a common and comparable scale with a mean of 0 and a standard deviation of 1, this approach does not resolve the alignment issue. Consider the scenario where a point $X_i$ from sequence $X$ has an identical value to $Y_j$ from sequence $Y$, yet the neighbourhood of $X_i$ is in a rising trend while the neighbourhood of $Y_j$ is in a falling trend. DTW may map these points onto one another to achieve minimal overall cost. 

To address this problem, ~\cite{Keogh2001ddtw} modifies DTW, denoted as DDTW. Instead of finding the optimal warping based on the raw values of the sequences, we consider the estimated local derivative of the sequence. The derivative of points on sequence $X$ is estimated by the following equations
\begin{equation}\label{formula:ddtw}
    \begin{aligned}
        &D_X[X_n] = \frac{(X_n - X_{n-1}) + ((X_{n+1}-X_{n-1})/2)}{2} \text{, with } n \in [2, \cdots, m - 1], \\
        &D_X[X_1] = D_X[X_2], \text{ and} \\
        &D_X[X_m] = D_X[X_m-1].
    \end{aligned}
\end{equation}
This is effectively the mean value between the slope of the line from the left neighbour to the point and the slope of the line from the left neighbour to the right neighbour. ~\cite{Keogh2001ddtw} suggests that after replacing the original sequence with the estimated derivative, the following procedure is the same as the classical dynamic time warping algorithm.

While DDTW considers the slope of the time series, it only considers the slope within a local neighbourhood, failing to consider the global features. An improvement was proposed in \cite{Zhao2016shapedtw} by dealing with multidimensional time series to account for both global features and local shapes, known as shape dynamic time warping (shapeDTW). The intuitive idea is to convert a one-dimensional time series \(X = (X_1, X_2, \ldots, X_m)\) of length \(m\) into a multidimensional series $\mathcal{D} = (d_1, \cdots, d_m) \in \mathbb{R}^{m \times l}$ , where each subsequence $d_i$ of length \(l\) embeds the information of the point \(X_i\). Each subsequence \( d_i \) can be constructed in one of two ways: either by directly taking the \( l \) values centred around the temporal point \( X_i \) to capture local shape information, or by applying the derivative operation (as in DDTW) to these values to emphasise local trends---this derivative-based approach is known as shapeDDTW. The dependent multidimensional DTW algorithm proposed in \cite{shokoohi2017generalizingdtw} is then applied to the two multidimensional series to calculate the distance cost and determine the optimal warping path. 

Following the convention in \cite{zhang2023dtw}, dynamic time warping can be used to detect the lag between two paths, $X$ and $Y$, by taking the mode of the differences between index pairs in the warping path \(\mathcal{W}\{X, Y\}\). Specifically, the lag is given by $\operatorname{Mode}(\Delta \mathcal{W}(X, Y))$, where $\Delta \mathcal{W}\{X, Y\} = \{\Delta w_1, \ldots, \Delta w_k\}$, and each \(\Delta w_k\) represents the difference between the indices $i$ and $j$ in the $k$-th warping pair, i.e., \(\Delta w_k = j - i\).

\section{From cross-sectional to network}\label{network_momentum_matrix}
We introduce a method to convert the pairwise lead-lag relationship into a `network momentum'. This network aims to capture the interconnected cross-sectional momentum between paired markets. The skew-symmetric lead-lag matrix, $\mathbf{V} \in \mathbb{R}^{M\times M}$, shows the directed lead-lag relationships in the market but has two limitations that prevent its direct use as network momentum.

First, the lead-lag matrix, obtained via dynamic time warping, contains only integer lags, i.e., $\mathbf{V}_{i,j} = \ell_{i,j} \in \mathbb{Z}$, which indicates that $\tilde{r}_{i,t-\ell_{i,j}}$ has some predictive power over $\tilde{r}_{j,t}$, but not its magnitude. While \cite{cartea2023detecting} suggests that the Lévy area can indicate the strength of a lead-lag relationship, both it and dynamic time warping face the second challenge: $\mathbf{V}$ is a dense matrix because every market $i$ has a lead-lag relationship with every other market $j$.

An ideal network momentum matrix should be sparse, retaining only the most significant connections and reducing noise. To address this, we propose fitting a graph learning model to the lead-lag matrix to generate a sparse adjacency matrix. This adjacency matrix keeps only the important edges, replacing the integer lag from dynamic time warping with non-negative weights that quantify the connection strength. We refer to this adjacency matrix as the network momentum matrix. Our method contrasts with the approach proposed by \cite{pu2023networkmomentumassetclasses}, where nodes in the graph are set with time-series momentum features, such as price information. In our case, each node represents a market and encodes its lead-lag relationships with other markets, reflecting the interconnected dynamics. We refer to this adjacency matrix as the network momentum matrix.

A non-parametric method to study the adjacency matrix is proposed in \cite[page 923]{kalofolias2016learngraphsmoothsignals}, where they study the following convex optimisation problem to learn the adjacency matrix.
\begin{definition}[Graph learning model]\label{definition:graph_learning_model}
    \cite[Page 923]{kalofolias2016learngraphsmoothsignals} Given a smooth matrix $\mathbf{X} \in \mathbb{R}^{N \times p}$ on a graph $G$, $\mathbf{D} \in \mathbb{R}_+^{N \times N}$ being the degree matrix of the graph, and sparsity parameters $\alpha > 0$ and $\beta \geq 0$, can be found by solving the following convex optimisation problem:
    \begin{equation*}
        \begin{aligned}
   & \operatorname{minimise}_{\mathbf{A}} \quad \operatorname{tr}(\mathbf{X}^T (\mathbf{D} - \mathbf{A}) \mathbf{X}) - \alpha \mathbf{1}^T \operatorname{log}(\mathbf{A}\mathbf{1}) + \beta \|\mathbf{A}\|^2_F \\
    & \text{such that} \quad \mathbf{A}_{i,j} = \mathbf{A}_{j,i}, \quad \mathbf{A}_{i,j} \geq 0 \quad \forall i \neq j, \quad \operatorname{diag}(\mathbf{A}) = \mathbf{0},
    \end{aligned}
    \end{equation*}
\end{definition}
Here, a logarithmic penalty term is used to prevent isolated nodes so that every node has at least one connected neighbour, and the Frobenius norm is used to control sparsity, unlike the $\ell_1$ norm, it only penalises the edge with large magnitude without disproportionately affecting smaller edges. This model also allows us to control the sparsity of the graph by tuning the parameters $\alpha$ and $\beta$, in general, as the parameters $\alpha$ and $\beta$ increase, a denser graph is obtained. 

The lead-lag matrix changes each time we fit the algorithm to new price data, so we denote the lead-lag matrix obtained at trading time $t$ as $\mathbf{V}_t$. We replace the signal matrix \(X\) in the graph learning model defined in \ref{definition:graph_learning_model} with \(\mathbf{V}_t\) to obtain an adjacency matrix with non-negative edge weights and no isolated markets. The edge values reflect the interconnected relationship of the leadingness of the markets.

It is suggested in \cite{pu2023networkmomentumassetclasses} that to mitigate the effects of scale differences in constructing network momentum—arising from the variance in the number of connections some nodes have, with some connected to many other assets and others to only a few—a graph normalisation should be applied to the adjacency matrix \(\mathbf{A}_t\) before using it to aggregate time series momentum. This normalisation is defined as follows:
\begin{equation}\label{formula:graph_normalisation}
    \tilde{\mathbf{A}}_t = \mathbf{D}_t^{-1/2} \mathbf{A}_t \mathbf{D}_t^{-1/2}
\end{equation}

The empirical analysis in \cite{pu2023networkmomentumassetclasses} suggests that combining $S$ adjacency matrices obtained from different lead-lag matrices \(\mathbf{V}_t\) based on historical price data with different lookback windows can improve performance. Therefore, we define the ensemble adjacency matrix \(\bar{\mathbf{A}}_t\) as:
\begin{equation}\label{formula:graph_ensemble}
    \bar{\mathbf{A}}_t = \frac{1}{S} \sum_{s=1}^S \mathbf{A}_t^{(s)}
\end{equation}
and will compare the strategy performance between using and not using the ensemble mechanism.

We now summarise the algorithm for calculating the network momentum matrix in Table \ref{alg:graph_learning}. In practice, the optimisation problem in the graph learning model \ref{definition:graph_learning_model} is solved numerically with MOSEK and Python library CVXPY \cite{diamond2016cvxpy}. 

\begin{algorithm}
\captionsetup{type=table}
\caption{Algorithm for Computing the Network Momentum Matrix Using Graph Learning}
\label{alg:graph_learning}
\begin{algorithmic}[1]
\REQUIRE Series of lead-lag matrices \(\{\mathbf{V}_t^s \in \mathbb{R}^{M \times M}\}_{s=1}^S\) observed at trading time \(t\), where \(M\) is the number of markets and \(S\) is the number of historical price data inputs, \(S \geq 1\).
\REQUIRE Hyperparameters \(\alpha > 0\) and \(\beta \geq 0\) for sparsity control.
\ENSURE Normalised network momentum matrix \(\tilde{\mathbf{A}}_t \in \mathbb{R}^{M \times M}\).
\STATE Initialise an ensemble adjacency matrix \(\bar{\mathbf{A}}_t\) with zeros of shape \((M, M)\).
\FOR{\(s = 1\) to \(S\)}
    \STATE Replace the signal matrix \(X\) in the graph learning model defined in \ref{definition:graph_learning_model} with \(\mathbf{V}_t^s\) to obtain the adjacency matrix \(\mathbf{A}_t^s\).
    \STATE Update the ensemble adjacency matrix according to \eqref{formula:graph_ensemble}: \(\bar{\mathbf{A}}_t \leftarrow \bar{\mathbf{A}}_t + \frac{1}{S} \mathbf{A}_t^s\).
\ENDFOR
\STATE Normalise \(\bar{\mathbf{A}}_t\) using the graph normalisation formula \eqref{formula:graph_normalisation} to obtain \(\tilde{\mathbf{A}}_t\).
\RETURN \(\tilde{\mathbf{A}}_t\)
\end{algorithmic}
\small{Note: If \(S=1\), this algorithm is equivalent to not using the graph ensemble method, thereby directly applying normalisation to the single obtained adjacency matrix.}
\end{algorithm}

\section{Methodology} \label{methodology}
\subsection{Data}
Our dataset contains the daily settlement price of 28 futures markets, ranging across sectors: agriculture, energy, metals and equity indices. The model training period spans from June 2002 to June 2024, with strategy performance evaluated on out-of-sample data from January 2005 to June 2024. 
A complete list of the markets included in our portfolio can be found in Appendix \ref{app:used_markets}.

We are interested in comparing different variations of our models for detecting network momentum, to each other, and to a baseline model. To facilitate this comparison we use the stationary block bootstrap procedure \cite{Politis1994stationarybootstrap} to generate a population set of price trajectories using the true market data. Stationary block bootstrapping preserves the auto- and cross-correlation of market returns, and so enables us to make statistical comparisons about the relative performance of the different models.

\subsection{Set Up for Time Series Momentum Features}\label{model_individual_mom}
In this section, we present the construction of classical time series momentum features based on price information. For each market $m$ and time index $t = 1, \ldots, T$, we denote the price for market $m$ at time $t$ as $P_{t,m} \in \mathbb{R}$, and therefore the price time series for market $m$ is denoted as 
$$
P_m := (P_{1,m}, P_{2,m}, \ldots, P_{T,m})\in \mathbb{R}^T.
$$
Note that futures markets are made up of a continually expiring sequence of individual contracts, and so there is not a continuously observed uninterrupted price for a futures market. We construct a continuous price using individual contract prices, according the the {\em backward Panama canal method} \cite{quantpediapanama}, and a suitable choice of roll dates. The derived price series is known as the {\em backadjusted price}. We construct $\mathbf{P} \in \mathbb{R}^{T \times M}$, representing a matrix of $M$ market backadjusted prices across a time horizon of $T$, where each vector is a price time series for a market. 

\begin{definition}[Price delta]\label{definition:price_return}
    Given a market $m$, we denote its price time series from time $t=1$ to $t=T$ as $(P_{1,m}, P_{2,m}, \ldots, P_{T,m})\in \mathbb{R}^T$, the price delta for it at time $t$, $\Delta_{t,m}$, is defined as the first difference in its price series,
    $$
    \Delta_{t,m} := P_{t,m} - P_{t-1,m}.
    $$
    Then, the price delta time series for market $m$ is denoted as $$\Delta_m := (\Delta_{1,m}, \ldots, \Delta_{T,m}),$$ and the matrix of market price deltas is defined as $\mathbf{\Delta} \in \mathbb{R}^{T\times M}$.
\end{definition}

Considering that each market exhibits different levels of price volatility, we choose to normalise the price deltas of each market to have unit volatility. This step aligns with the extant literature \cite{pu2023networkmomentumassetclasses, Baz2015DissectingIS} in their construction of time series momentum features. 

\begin{definition}[Volatility-scaled price delta]\label{definition:vs_price_delta}
    Given a market $m$, denote its price delta time series from time $t=1$ to $t=T$ as $(\Delta_{1,m}, \ldots, \Delta_{T,m})$, let the exponential weighted moving standard deviation at time $t$ over a span of 22 days denoted as $\sigma_{t,m}^{22}$. The volatility scaled price deltas for market $m$ at time $t$ is defined as 
    $$
    \tilde \Delta_{t,m} := \frac{\Delta_{t,m}}{\sigma_{t,m}^{22}}.
    $$
    The time series of volatility-scaled price delta for market $m$ is denoted as $$\tilde \Delta_m =(\tilde \Delta_{1,m}, \ldots, \tilde \Delta_{T,m}),$$ and the matrix of all volatility-scaled market price deltas is defined as $\mathbf{\tilde \Delta}\in \mathbb{R}^{T\times M}$.
\end{definition}

\begin{definition}[Volatility-scaled price]\label{definition:vs_price}
    Given a market $m$, denote its volatility-scaled price delta time series from time $t=1$ to $t=T$ as $(\tilde \Delta_{1,m}, \ldots, \tilde \Delta_{T,m})$, the volatility-scaled price for market $m$ at time $t$ is defined as
    $$
    \tilde P_{t,m} := \sum_{i=0}^{t} \tilde \Delta_{i,m}.
    $$
    The time series of volatility-scaled price for market $m$ is denoted as $$\tilde P := (\tilde P_{1,m}, \ldots, \tilde P_{T,m}),$$ and the matrix of all market price deltas is defined as $\mathbf{\tilde P}\in \mathbb{R}^{T\times M}$.
\end{definition}

Now, we are ready to define the time series momentum features, which we also refer to as oscillators or individual momentum features interchangeably. 

\begin{definition}[Time series momentum features]\label{definition:oscillator}
    Given a volatility-scaled price time series for market $m$ from time $t=1$ to $t=T$, $\tilde P_m = (\tilde P_{1,m}, \ldots, \tilde P_{T,m})$, and a speed parameter $k \in \mathbb{N}_+$, we define two smoothing factors as,
    $$
    \alpha_{fast}(k) := \frac{1}{2^k}, \quad \alpha_{slow}(k) := \frac{1}{M\times 2^k}.
    $$
    for some $M > 1$. Let $\mu(P, \alpha)$ denote the exponential-weighted moving average of time-series $P$, with decay factor $\alpha$. Then the oscillator for market $m$ and speed $k$, denoted $R_{t,m}^k$, is defined by:
    $$
    R_{t,m}^k := \mu(\tilde{P}_m, \alpha_{fast}(k)) - \mu(\tilde{P}_m, \alpha_{slow}(k))
    $$
\end{definition}
The underlying intuition is that the crossover of exponentially weighted moving averages can provide insight into recent market trends. If the short-term average price crosses above the long-term average price from below, it indicates an expected increase in price, suggesting a potential upward trend and, conversely, a downward trend if it crosses below. Also, for smaller speed parameters $k$, the time series momentum feature contains information for short-term recent trends. Conversely, time series momentum features with larger speed parameter $k$ contain long-term trend information. Throughout this paper, we choose $k = \{1,2,3,4,5,6\}$ to create $6$ time series momentum features at different speeds to identify the auto-correlation in each market, in line with existing literature \cite{Baz2015DissectingIS, pu2023networkmomentumassetclasses, lim2020enhancingtimeseriesmomentum}.

\subsection{Set Up for Network Momentum Features}\label{model_network_mom}
Now we introduce the setup for the network momentum features. At each trading day $t$, given a lookback window of $\delta$ days, the first step of constructing network momentum features is to use one of the lead-lag detection methods (L\'evy Area, or one- or multi-dimensional DTW) to construct a lead-lag matrix $\mathbf{V}_t$ by using the volatility-scaled prices deltas $\mathbf{\tilde \Delta}_t \in \mathbb{R}^{\delta \times M}$ as input features. Each vector in $\mathbf{\tilde \Delta}_t$ is the volatility-scaled price delta for a market across the past $\delta$ days from day $t$. The second step is to apply the graph learning algorithm in Table \ref{alg:graph_learning} by using the lead-lag matrix $\mathbf{V}_t$ as an input feature to obtain the normalised adjacency matrix $\mathbf{\tilde A}_t$.

\begin{definition}[Network momentum feature]\label{definition:network_feature}
    Given a time series momentum feature with speed $k$ for market $m$ at time $t$, $R_{t,m}^k$, and the normalised adjacency matrix $\mathbf{\tilde A}_t$ from fitting a lead-lag detection model and the graph learning model to the volatility-scaled price deltas $\mathbf{\tilde \Delta}_t$, the network momentum feature with speed $k$ for market $m$ at time $t$ is defined as
    $$
    \tilde{R}_{t,m}^k := \sum_{n \in \mathcal{N}_t(m)} \tilde{A}_{m,n} R_{t,m}^k,
    $$
    where \(\mathcal{N}_t(m)\) denotes the set of markets connected to market \(m\) such that \(\tilde{A}_{m,n} \neq 0\) and $R_{t,n}^k$ is the time series momentum feature with the same speed for market $n$ at time $t$.
\end{definition}

Considering that the graph sparsity is significantly influenced by the two hyperparameters $\alpha$ and $\beta$ in the graph learning model \ref{definition:graph_learning_model}, which consequently affect the number of connections each market can establish, we conduct a discrete grid search over the combinations of:
\begin{equation*}
    \alpha = \{0.001, 0.01, 0.1, 1, 10, 100\}\text{,} \quad \beta = \{0.001, 0.01, 0.1, 1, 10, 100\},
\end{equation*}
on in-sample data to determine their optimal combination for achieving the highest net Sharpe ratio.

We choose $\delta=132$, so the lead-lag matrix is constructed by considering each market's past half year's daily performances. In addition, to enhance the robustness of our model, we consider employing an ensemble method that fits multiple lead-lag matrices to $\mathbf{\tilde \Delta}_t$ across a range of lookback windows. Specifically, we use the following series of lookback windows:
$$
\delta = \{22, 44, 66, 88, 110, 132\}.
$$
The multiple lead-lag matrices are summarised into a series, which serves as a new input to the graph learning algorithm detailed in Table \ref{alg:graph_learning}. According to \cite{pu2023networkmomentumassetclasses}, employing an ensemble method helps reduce the variance of the learned edge weights, improving the strategy's performance and reducing turnover.

\section{Portfolio Construction}\label{portfolio construction}
In this section we detail our portfolio construction methodology and baseline model. We then introduce our variations of network momentum and compare them to the baseline. The raw inputs to the portfolio are the moving-average crossover signals for each market, at various speeds, described in the previous section. The first step in the portfolio construction is to attenuate these raw signals by passing them through a {\em response function} which has the effect of exponentially squashing the signal towards zero in its left and right tails. If the raw signal sign is an indication of the strength of a trend (both upwards and downwards), then the response function serves the purpose of risk control, in the sense that it curtails the response to extreme trends, reflecting increasing uncertainty about the predictive power of the trend signal at its extremes. To this end, we choose a {\em reverting sigmoid} function \cite{martin2023design} with the following functional form:

\begin{definition}[Response function]\label{formula:response_function}
    The response function \( r(x): \mathbb{R} \rightarrow \mathbb{R} \), parameterised by a positive constant $\lambda >0$, is defined as follows:
    $$
    r(x) := c_\lambda \cdot x \cdot e^{-\lambda^2 x^2 / 2},
    $$
\end{definition}
In accordance with our desire to combine signals with unit variance, the response function includes the term $c_\lambda$ which acts as a normalisation constant. The response function peaks at \( x = \pm \lambda^{-1} \), setting the bounds for the maximum response to the trend signal; we fix $\lambda = \sqrt{2}$.

\begin{definition}[Position signal]\label{definition:position_signal}
    Given a series of momentum features with different speeds for market $m$ at time $t$, $(R_{t,m}^1, \ldots, R_{t,m}^K)$, the desired daily position, in lots, for market $m$ at time $t$, $X_{t, m}$ is defined as:
    $$
    X_{t,m} := \frac{1}{M}\left(\frac{1}{K}\sum_{k=1}^K r(R_{t,m}^k) \right) \cdot (F_{t,m} \cdot E_{t,m} \cdot \sigma_{t,m}^{22})^{-1} \cdot \Gamma \cdot \frac{\sigma_{\text{tgt}}}{\sqrt{252}},
    $$
    where
    \begin{itemize}
        \item $F_{t,m}$ denotes the {\em point value} of the futures contract: the local currency value of a 1 point move in the price of the contract,
        \item $E_{t,m}$ denotes the exchange rate between the currency in which market $m$ trades and the USD on day $t$,
        \item $\sigma_{t,m}^{22}$ is the exponential weighted
moving standard deviation of the price delta of market $m$ at time t over a span of 22 days,
        \item $\Gamma$ denotes the {\em notional aum} allocated to the portfolio, in USD,
        \item $\sigma_{\text{tgt}}$ denotes the annual target portfolio volatility which we set at $10\%$,
        \item $r(\cdot)$ is the response function in Definition \ref{formula:response_function}.
    \end{itemize}
\end{definition}

In this construction, we take equal contributions from oscillators with different speeds and markets in our portfolio, accounted for by the scalars $\frac{1}{M}$ and $\frac{1}{K}$ respectively. The term $(F_{t,m} \cdot E_{t,m} \cdot \sigma_{t,m}^{22})^{-1}$ is the number of lots of market $m$ on day $t$ required to achieve 1 USD of risk. The last part in the above definition, $\left(\Gamma \cdot \frac{\sigma_{\text{tgt}}}{\sqrt{252}}\right)$, is used to scale our position to realise the daily USD risk amount that our portfolio is targeting. The desired daily position, as defined above, constitutes our baseline portfolio. In practice, the weights controlling the contribution of the oscillators, and the participation of each market in the portfolio, would be subject to selection via an optimisation process, where the objective function would attempt to maximise some desirable portfolio metric (typically the net Sharpe ratio). It is our intention to compare the relative merits of various network momentum indicators to a baseline trend-following portfolio, so we choose to work with an unoptimised (at least with respect to oscillator and market weights) portfolio. Given the baseline portfolio above, the portfolios constructed using the network momentum signals for each market $m$ at time $t$, $\tilde X_{t,m}$, are defined similarly by replacing the momentum features in Definition \ref{definition:position_signal} with the network momentum features in Definition \ref{definition:network_feature}.

The daily gross USD pnl of market $m$ generated by the position signal $X_{t,m}$ for time series momentum features is calculated as
$$
r_{t+2,m} := X_{t,m} \cdot \Delta_{t+2,m} \cdot F_{t+2,m} \cdot E_{t+2, m}.
$$and the daily gross USD pnl for network momentum features $\tilde r_{t+2,m}$ is calculated similarly by replacing $X_{t,m}$ with $\tilde X_{t,m}$. We take a conservative approach to pnl calculation, assuming that a signal/position generated at time $t$, including information up to time $t$, is established via a trade at time $t+1$, and then pnl is earned on the position at time $t+2$. This is in contrast to examples in the literature that assume pnl from a signal at time $t$ is earned in full on the position at time $t+1$, but this is unrealistic, unless the position is established at or close to the opening price on day $t+1$, and may also be compromised by market synchronicity issues. To calculate the net USD pnl, we need to incorporate transaction costs, which typically are a market specific cost to establishing each new daily desired position. We choose market specific transaction costs based on half the average bid-ask spread in the market, denoted $s_m$, over some representative historical period. The transaction cost for establishing the position $X_{t,m}$ on day $t+1$ is then calculated as:
$$
c_{t+1,m} := \left| X_{t+1,m} - X_{t,m} \right| \cdot \frac{s_{m}}{2} \cdot F_{t+1,m} \cdot E_{t+1,m}.
$$
Here we estimate the cost for executing the trade by half of the spread $s_{t,m}$, and this execution happens during day $t+1$. We denote the transaction cost for network momentum features as $\tilde c_{t+1,m}$ by replacing $X_{t,m}$ with $\tilde X_{t,m}$.
Finally, we can calculate the net daily USD pnl for market $m$ on time $t$ by
$$
r^{\prime}_{t,m} = r_{t,m} - c_{t,m}.
$$
This formulation means that the net pnl generated by market \(m\) on day \(t\) consists of the gross return realised from position \(X_{t-2, m}\) combined with the transaction cost incurred for establishing the new position generated by \(X_{t-1,m}\). The net pnl for network momentum features is denoted as $\tilde r^{\prime}_{t,m}$ by replacing $X_{t,m}$ with $\tilde X_{t,m}$. We can transform the daily pnl into a return by dividing through by the notional aum, $\Gamma$.

We utilise a uniform methodology for portfolio construction and returns calculation across a series of candidate network momentum models, each employing different lead-lag detection algorithms. The models and their configurations are defined as follows:

\begin{enumerate}
    \item \textbf{MACD} uses the time series momentum signal \( R_m^k \), for \( k \) from 1 to 6, as defined in Definition \ref{definition:oscillator}, to calculate the position signal in Definition \ref{definition:position_signal}; this is our baseline model portfolio.
    
    \item \textbf{NMM-DTW} and \textbf{NMM-DTW-E}:
    \begin{itemize}
        \item At each training time \(t\), NMM-DTW constructs the lead-lag matrix \(\mathbf{V}_t\) using the classical dynamic time warping algorithm with a \(\delta=132\) lookback window. This matrix inputs into the graph learning model and computes the normalised adjacency matrix \(\mathbf{\tilde A}_t\) following the algorithm in Table~\ref{alg:graph_learning}. The network momentum features derived from \(\mathbf{\tilde A}_t\) following Definition~\ref{definition:network_feature} are then used to calculate the position signal (Definition~\ref{definition:position_signal}).

        \item NMM-DTW-E employs ensemble methods by fitting the DTW algorithm to varying lookback windows \(\delta = \{22, 44, 66, 88, 110, 132\}\). The resulting series of lead-lag matrices serves as the new input to the graph learning model. Subsequent steps follow the NMM-DTW process.
    \end{itemize}
    
    \item \textbf{NMM-DDTW} and \textbf{NMM-DDTW-E}:
    \begin{itemize}
        \item NMM-DDTW uses the derivative dynamic time warping algorithm. Subsequent steps follow the NMM-DTW process.
        \item NMM-DDTW-E employs the ensemble methods by constructing the lead-lag matrices with DDTW algorithm with lookback \(\delta = \{22, 44, 66, 88, 110, 132\}\). Subsequent steps follow the NMM-DTW-E process.
    \end{itemize}
    
    \item \textbf{NMM-SDTW} and \textbf{NMM-SDTW-E}:
    \begin{itemize}
        \item NMM-SDTW applies the shape dynamic time warping algorithm, with the descriptor length $l=11$, to construct the lead-lag matrix with a lookback window of $\delta=132$. Subsequent steps follow the NMM-DTW process.
        \item NMM-SDTW-E employs the ensemble methods with shape dynamic time warping with lookback windows \(\delta = \{22, 44, 66, 88, 110, 132\}\). Subsequent steps follow the NMM-DTW-E process.
    \end{itemize}

    \item \textbf{NMM-SDDTW} and \textbf{NMM-SDDTW-E} :
    \begin{itemize}
        \item NMM-SDDTW applies the shape dynamic time warping derivative algorithms, with the descriptor length $l=11$, to construct the lead-lag matrix with a lookback window of $\delta=132$. Subsequent steps follow the NMM-DTW process.
        \item NMM-SDTW-E employs the ensemble methods with lookback windows \(\delta = \{22, 44, 66, 88, 110, 132\}\). Subsequent steps follow the NMM-DTW-E process.
    \end{itemize}
    
    \item \textbf{NMM-LEVY} and \textbf{NMM-LEVY-E}: 
    \begin{itemize}
        \item NMM-LEVY applies the L\'evy area algorithm to construct the lead-lag matrix with lookback window \(\delta=132\). Subsequent steps follow the NMM-DTW process.
        \item NMM-LEVY-E employs the ensemble methods with L\'evy area algrotim with lookback windows \(\delta = \{22, 44, 66, 88, 110, 132\}\). Subsequent steps follow the NMM-DTW-E process.
    \end{itemize}
\end{enumerate}

\section{Performance Analysis}\label{analysis}
\subsection{Portfolio Performance Analysis}
In evaluating model performance, we consider the following three aspects, following the convention established in \cite{pu2023networkmomentumassetclasses}:
\begin{enumerate}
    \item \textbf{Profitability}: This includes annualised expected gross return, annualised expected net return, and hit rate -- defined as the percentage of days with positive returns during the out-of-sample periods.
    \item \textbf{Risk}: This includes volatility, downside deviation, and maximum drawdown to understand the risk exposure of our models. While comparing the volatility and downside deviation across models is unnecessary due to each model's positions being scaled to a target volatility of 10\% (Definition \ref{definition:position_signal}), we include them in our summary for completeness, as they remain relevant for calculating the Sharpe ratio and Sortino ratio.
    \item \textbf{Overall and Other Performance}: This includes transaction costs, skewness of monthly returns, Sharpe ratio (expected return / volatility), Sortino ratio (expected return / downside deviation), Calmar ratio (expected return / maximum drawdown), and the ratio of average profits to average losses \(\left(\frac{\text{Avg. P}}{\text{Avg. L}}\right)\).
\end{enumerate}

We present the performance of the benchmark MACD model alongside the network momentum models. Our primary focus is on their average performance across $100$ bootstrapped datasets. We investigate whether the network momentum models can achieve a statistically significant higher net Sharpe ratio compared to the benchmark MACD model. We also present the models' performance on real-world price data from the out-of-sample period of 2005 to 2024 to illustrate their profitability in a historical context.

\begin{table}[htbp]
\centering
\caption{Performance Metrics for Various Signals}
\label{tab:overall_performance}
\resizebox{\textwidth}{!}{%
\begin{threeparttable}
\begin{tabular}{lcccccccccccc}
\hline
& Gross & Transaction & Net & vol. & Sharpe & downside & MDD & Sortino & Calmar  & Skewness & hit rate & Avg. P  \\
& Return & & Return & & &deviation & & & & & & Avg. L  \\
\hline
\multicolumn{11}{c}{\textbf{Panel A: Average Performance on 100 Bootstrapped Price Data}} \\
\hline
MACD & 0.057 & 0.027 & 0.030 & 0.107 & 0.277 & 0.058 & 0.239 & 0.515 & 0.039 & 0.395 & 0.516 & 1.158  \\
NMM-DTW & 0.064 & 0.029 & 0.034 & 0.108 & 0.315 & 0.058 & 0.261 & 0.592 & 0.041 & 0.441 & 0.515 & 1.200  \\
NMM-DTW-E & 0.063 & 0.023 & 0.039 & 0.109 & 0.353 & 0.058 & 0.259 & 0.669 & 0.046 & 0.457 & 0.516 & 1.226  \\
NMM-DDTW & \textbf{0.064} & 0.029 & 0.034 & 0.109 & 0.315 & 0.059 & 0.256 & 0.590 & 0.042 & 0.450 & 0.514 & 1.203  \\
NMM-DDTW-E & 0.063 & 0.023 & \textbf{0.039} & 0.110 & \textbf{0.357} & 0.058 & 0.250 & \textbf{0.684} & 0.048 & 0.486 & 0.514 & 1.243  \\
NMM-SDTW & 0.064 & 0.028 & 0.035 & 0.110 & 0.319 & 0.058 & 0.280 & 0.606 & 0.041 & 0.458 & 0.508 & 1.235  \\
NMM-SDTW-E & 0.062 & \textbf{0.022} & 0.039 & 0.109 & 0.355 & 0.058 & 0.255 & 0.677 & 0.047 & 0.473 & 0.512 & 1.250  \\
NMM-SDDTW & 0.062 & 0.029 & 0.032 & 0.109 & 0.296 & 0.057 & 0.269 & 0.568 & 0.039 & 0.507 & 0.504 & 1.234 \\
NMM-SDDTW-E & 0.062 & 0.023 & 0.038 & 0.110 & 0.350 & 0.057 & 0.257 & 0.679 & 0.046 & \textbf{0.509} & 0.510 & \textbf{1.255} \\
NMM-LEVY & 0.064 & 0.027 & 0.036 & 0.109 & 0.336 & 0.059 & \textbf{0.230} & 0.624 & \textbf{0.050} & 0.419 & \textbf{0.517} & 1.206 \\
NMM-LEVY-E & 0.060 & 0.024 & 0.035 & 0.109 & 0.323 & 0.058 & 0.240 & 0.610 & 0.045 & 0.454 & 0.516 & 1.202 \\
\hline
\multicolumn{11}{c}{\textbf{Panel B: Performance on Real Price Data}} \\
\hline 
MACD & 0.051 & 0.026 & 0.024 & 0.104 & 0.233 & 0.053 & 0.227 & 0.454 & 0.031 & 0.645 & \textbf{0.526} & 1.080 \\
NMM-DTW & 0.054 & 0.028 & 0.026 & 0.106 & 0.243 & 0.056 & 0.274 & 0.457 & 0.027 & 0.630 & 0.513 & 1.147 \\
NMM-DTW-E & 0.062 & 0.023 & \textbf{0.039} & 0.106 & \textbf{0.364} & 0.056 & 0.203 & 0.694 & 0.055 & 0.683 & 0.509 & 1.285 \\
NMM-DDTW & 0.056 & 0.028 & 0.027 & 0.107 & 0.257 & 0.053 & 0.247 & 0.517 & 0.032 & \textbf{0.759} & 0.487 & 1.282 \\
NMM-DDTW-E & 0.055 & 0.022 & 0.032 & 0.107 & 0.298 & 0.055 & 0.244 & 0.577 & 0.038 & 0.719 & 0.513 & 1.198 \\
NMM-SDTW & \textbf{0.065} & 0.027 & 0.037 & 0.106 & 0.351 & 0.054 & \textbf{0.163} & 0.689 & \textbf{0.066} & 0.704 & 0.513 & 1.249 \\
NMM-SDTW-E & 0.055 & 0.022 & 0.033 & 0.107 & 0.307 & 0.055 & 0.222 & 0.600 & 0.043 & 0.691 & 0.513 & 1.205 \\
NMM-SDDTW & 0.049 & 0.028 & 0.020 & 0.108 & 0.189 & 0.058 & 0.289 & 0.354 & 0.020 & 0.589 & 0.470 & \textbf{1.303} \\
NMM-SDDTW-E & 0.057 & \textbf{0.022} & 0.035 & 0.106 & 0.328 & 0.054 & 0.208 & 0.643 & 0.048 & 0.720 & 0.509 & 1.249 \\
NMM-LEVY & 0.064 & 0.026 & 0.038 & 0.106 & 0.356 & 0.054 & 0.204 & \textbf{0.702} & 0.053 & 0.675 & 0.517 & 1.228 \\
NMM-LEVY-E & 0.055 & 0.023 & 0.032 & 0.106 & 0.300 & 0.054 & 0.208 & 0.586 & 0.044 & 0.673 & 0.513 & 1.198 \\
\hline
\end{tabular}
\begin{tablenotes}[para,flushleft]
\small
\item[a] Best performance is in bold. 
\item[b] No comparison for volatility and downside deviation since every portfolio is scaled to the same target annualised volatility in \ref{definition:position_signal} for direct comparison of the net Sharpe.
\end{tablenotes}
\end{threeparttable}}
\end{table}

In Panel A of Table \ref{tab:overall_performance}, we report the average performance of the portfolio constructed from various momentum models on bootstrapped price data. In Panel B, we report the performance of these models on real price data from 2005 to 2024. 

Based on the metrics in Panel A, all network momentum models (NMM) exhibit higher expected gross returns than the benchmark MACD model, with the NMM-DDTW model achieving the highest at $0.064$, compared to MACD's $0.057$. Typically, NMM models incur higher transaction costs than MACD, reflecting their sensitivity to market movements and increased daily turnover. However, ensemble methods reduce transaction costs, with DTW variations further decreasing them to $0.022$, approximately 19\% lower than MACD. The NMM-LEVY model achieves an 11\% reduction in costs. As a result, all NMM models demonstrate better performance over MACD in terms of expected net returns, net Sharpe ratios, and Sortino ratios. Notably, NMM-DDTW-E achieves a Sharpe ratio of $0.357$ and a Sortino ratio of $0.684$, marking increases of 29\% and 33\%, respectively, over MACD.

The ability to effectively follow trends is crucial for trading strategies. NMM-SDDTW-E stands out with the highest \(\frac{\text{Avg. Profit}}{\text{Avg. Loss}}\) ratio among all NMM models. It also exhibits the highest positive skewness, suggesting that although it may frequently incur small losses, the gains it captures are significant. Meanwhile, NMM-LEVY demonstrates the smallest MDD and highest hit rate, suggesting it is particularly effective at identifying trend reversals and capturing new opportunities for positive returns, although it achieves smaller gains per trade, as indicated by its slightly lower skewness and the ratio between average profit and average loss.

Panel B of Table \ref{tab:overall_performance} demonstrates that NMM models outperform the benchmark MACD model on real market data during the out-of-sample period from 2005 to 2024. NMM-DTW-E achieves the highest net Sharpe ratio at $0.364$, compared to the benchmark's $0.233$, showing better risk-adjusted returns. NMM-DDTW exhibits the most positive skew in returns at $0.759$, surpassing the benchmark's $0.645$. Although MACD has the highest hit rate, indicating more days with positive PnL, it suffers from the lowest \(\frac{\text{Avg. Profit}}{\text{Avg. Loss}}\) ratio, suggesting that its losses are larger than those of NMM models. However, we reiterate that the portfolio included in Appendix \ref{app:used_markets} is somewhat random, and the models' performance may not be reproducible for other portfolios. Therefore, we emphasise that our assessment of the models is primarily based on their performance on the bootstrapped data.

\begin{figure}[htbp]
    \centering
    \includegraphics[width=.9\textwidth]{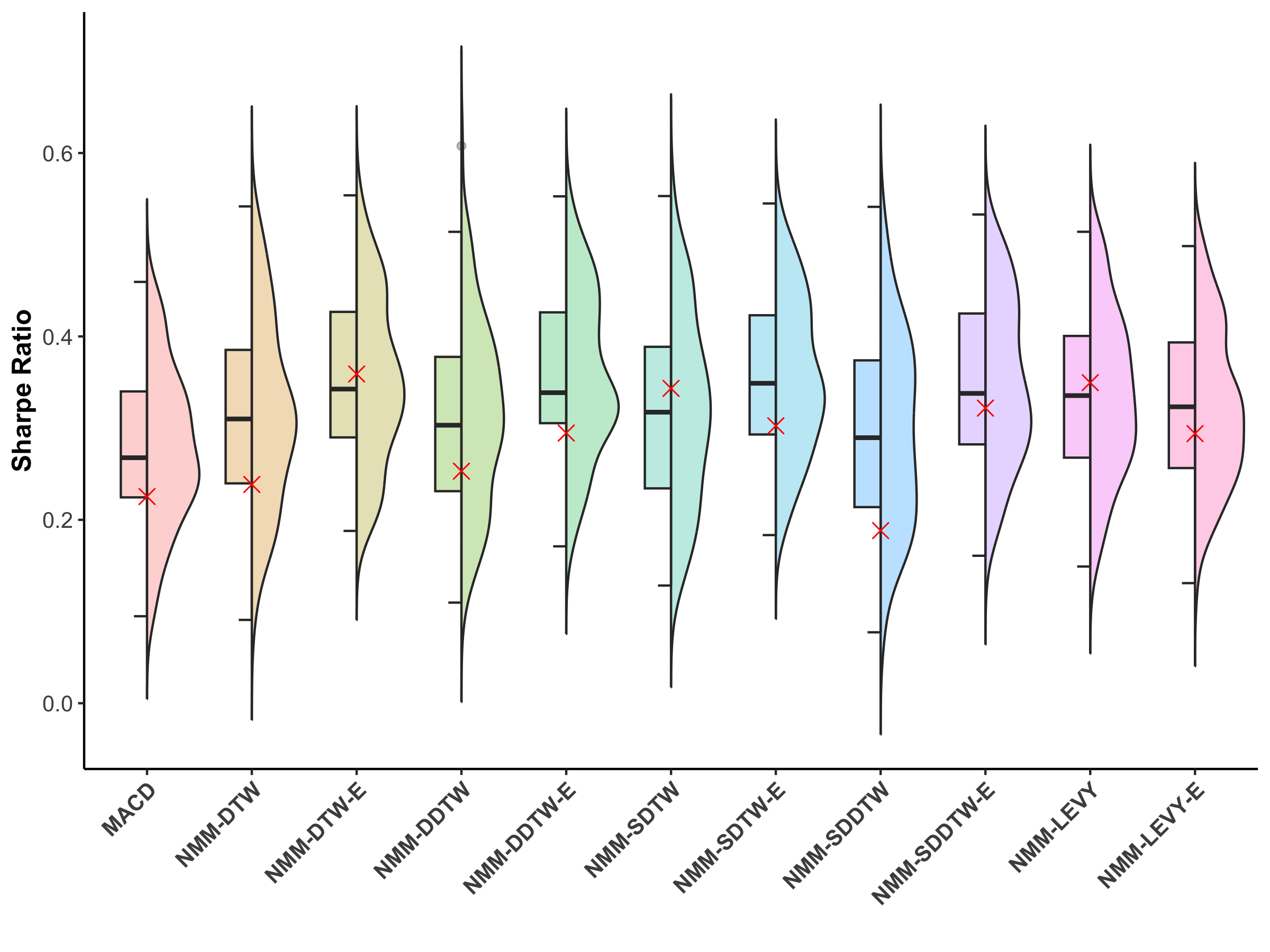}
    \caption{Distribution of net Sharpe Ratios for the Benchmark Model (MACD) and Network Momentum Models on bootstrapped datasets, with net Sharpe achieved on real price data indicated by red crosses}
    \label{fig:sharpe_box}
\end{figure}

We examine the distribution of the net Sharpe ratios for all models. Figure~\ref{fig:sharpe_box} presents the distribution of Sharpe ratios on the bootstrapped price datasets along with their interquartile ranges. The net Sharpe ratios achieved by each model on the actual price data are marked by red crosses on the distribution plots. The box plots demonstrate that the median net Sharpe ratios for all network momentum models are higher than those for the MACD model, and the ensemble methods further enhance performance. The positioning of the red crosses, which for all models except NMM-DDTW-E and NMM-SDTW-D fall within the interquartile ranges, suggests that the bootstrapped price data provides a valid representation of the real price data and is suitable for inference.

We have two primary objectives as follows:
\begin{enumerate}
    \item To determine whether the net Sharpe ratio achieved by the network momentum model is significantly higher than that achieved by the MACD model when both are used to construct portfolios from the same price data set. We employ a one-sided Wilcoxon signed-rank test~\cite{wilcoxon1992individual}, a matched-pair test, to assess if the difference in net Sharpe ratios (network momentum model minus MACD model) is significantly greater than 0.
    \item To examine whether the distributions of the net Sharpe ratios from the MACD model and a network momentum model are statistically different without considering the matched-pair nature of the data. We use the one-sided Kolmogorov-Smirnov test~\cite{berger2014kolmogorov} to determine if the cumulative distribution function of the MACD model's net Sharpe ratios is stochastically greater than that of the network momentum model, it indicates that the MACD model generally yields lower Sharpe ratios than the network momentum model.
\end{enumerate}

We report the p-values for the two tests in Table~\ref{tab:pvalues}. For the Wilcoxon signed-rank test, all network momentum models achieve significant p-values ($p < 0.05$). This demonstrates that, when applied to the same random set of market price data, the network momentum models significantly outperform the benchmark MACD model, which relies only on time-series momentum in terms of net Sharpe ratio. For the Kolmogorov-Smirnov test, aside from NMM-SDDTW, all other NMM models achieve significant p-values ($p < 0.05$). This indicates that the cumulative distribution function of the net Sharpe ratios for the network momentum models is stochastically smaller than that of the MACD model, suggesting that the network momentum models generally achieve higher Sharpe ratios than the MACD model. These two tests collectively underscore the enhanced performance capability of the network momentum feature.

Our results demonstrate the robustness and reliability of the network momentum spillover identified by the proposed algorithms. These findings suggest that under both uniform and varied market conditions, the NMM models consistently outperform the benchmark MACD model with statistically confidence.

\begin{table}[htbp]
\centering
\caption{P-Values for Sharpe Ratio Comparisons Against Benchmark}
\label{tab:pvalues}
\resizebox{\textwidth}{!}{%
\begin{tabular}{lcccccccccc}
\hline
& NMM-DTW& NMM-DTW-E & NMM-DDTW& NMM-DDTW-E& NMM-SDTW& NMM-SDTW-E& NMM-SDDTW& NMM-SDDTW-E& NMM-LEVY& NMM-LEVY-E\\
\hline
Wilcoxon signed-rank test & 0 & 0 & 0 & 0 & 0 & 0 & 0.005 & 0 & 0 & 0 \\
Kolmogorov–Smirnov test & 0.018 & 0 & 0.012 & 0 & 0.005 & 0 & 0.077 & 0 & 0 & 0.002 \\
\hline
\end{tabular}}
\end{table}

\subsection{Long/Short Performance Analysis}
In this section, We focus on the model's ability to identify and respond to upward and downward market trends by examining performance in both long and short trading positions. The returns from these positions are analysed separately, with the metrics for short and long positions detailed in Tables \ref{tab:short_performance} and \ref{tab:long_performance}, respectively.

\begin{table}[htbp]
\centering
\caption{Performance Metrics for Various Signals in Short Direction Only}
\label{tab:short_performance}
\resizebox{\textwidth}{!}{%
\begin{threeparttable}
\begin{tabular}{lcccccccccccc}
\hline
& Gross & Transaction & Net & vol. & Sharpe & downside & MDD & Sortino & Calmar  & Skewness & hit rate & Avg. P  \\
& Return & & Return & & &deviation & & & & & & Avg. L  \\
\hline
\multicolumn{11}{c}{\textbf{Panel A: Average Performance on 100 Bootstrapped Price Data}} \\
\hline
MACD & -0.011 & 0.014 & -0.026 & 0.066 & -0.396 & 0.040 & 0.635 & -0.638 & -0.012 & 0.804 & \textbf{0.367} & 1.245  \\
NMM-DTW & \textbf{-0.005} & 0.014 & -0.020 & 0.062 & -0.329 & 0.039 & 0.546 & -0.513 & -0.010 & 0.885 & 0.351 & 1.379  \\
NMM-DTW-E & -0.006 & 0.012 & -0.018 & 0.058 & -0.317 & 0.037 & 0.519 & -0.490 & -0.010 & 1.007 & 0.342 & 1.427  \\
NMM-DDTW & -0.006 & 0.014 & -0.021 & 0.062 & -0.340 & 0.038 & 0.553 & -0.541 & -0.011 & 1.042 & 0.349 & 1.385  \\
NMM-DDTW-E & -0.005 & 0.011 & \textbf{-0.017} & 0.059 & \textbf{-0.300} & 0.037 & 0.508 & \textbf{-0.467} & \textbf{-0.010} & \textbf{1.155} & 0.341 & 1.449  \\
NMM-SDTW & -0.007 & 0.014 & -0.022 & 0.061 & -0.364 & 0.038 & 0.562 & -0.569 & -0.011 & 0.996 & 0.348 & 1.358  \\
NMM-SDTW-E & -0.006 & 0.011 & -0.017 & 0.058 & -0.310 & 0.037 & \textbf{0.500} & -0.474 & -0.010 & 1.035 & 0.342 & 1.439  \\
NMM-SDDTW & -0.007 & 0.014 & -0.021 & 0.061 & -0.351 & 0.039 & 0.555 & -0.544 & -0.011 & 0.877 & 0.344 & 1.395  \\
NMM-SDDTW-E & -0.007 & \textbf{0.011} & -0.018 & 0.058 & -0.317 & 0.037 & 0.510 & -0.487 & -0.010 & 1.130 & 0.333 & \textbf{1.468} \\
NMM-LEVY & -0.008 & 0.014 & -0.022 & 0.062 & -0.363 & 0.039 & 0.573 & -0.576 & -0.011 & 0.814 & 0.357 & 1.327 \\
NMM-LEVY-E & -0.010 & 0.012 & -0.022 & 0.060 & -0.374 & 0.038 & 0.575 & -0.586 & -0.011 & 0.861 & 0.341 & 1.387  \\
\hline
\multicolumn{11}{c}{\textbf{Panel B: Performance on Real Price Data}} \\
\hline 
MACD & -0.013 & 0.014 & -0.028 & 0.070 & -0.396 & 0.043 & 0.584 & -0.645 & -0.014 & 0.953 & \textbf{0.376} & 1.189  \\
NMM-DTW & -0.007 & 0.014 & -0.022 & 0.066 & -0.327 & 0.040 & 0.525 & -0.536 & -0.012 & 1.218 & 0.342 & 1.430  \\
NMM-DTW-E & \textbf{-0.005} & 0.012 & \textbf{-0.016} & 0.064 & -0.254 & 0.039 & \textbf{0.450} & \textbf{-0.410} & \textbf{-0.010} & 1.237 & 0.363 & 1.368  \\
NMM-DDTW & -0.009 & 0.015 & -0.023 & 0.066 & -0.353 & 0.040 & 0.514 & -0.578 & -0.013 & 1.295 & 0.368 & 1.252  \\
NMM-DDTW-E & -0.007 & 0.012 & -0.019 & 0.063 & -0.299 & 0.039 & 0.487 & -0.484 & -0.011 & \textbf{1.395} & 0.333 & 1.485  \\
NMM-SDTW & -0.006 & 0.014 & -0.020 & 0.067 & -0.297 & 0.041 & 0.456 & -0.483 & -0.013 & 1.243 & 0.350 & 1.393  \\
NMM-SDTW-E & -0.007 & 0.012 & -0.018 & 0.062 & \textbf{-0.297} & 0.040 & 0.485 & -0.465 & -0.011 & 1.290 & 0.355 & 1.350  \\
NMM-SDDTW & -0.012 & 0.014 & -0.025 & 0.067 & -0.381 & 0.041 & 0.553 & -0.615 & -0.013 & 1.164 & 0.312 & \textbf{1.575} \\
NMM-SDDTW-E & -0.008 & \textbf{0.012} & -0.019 & 0.062 & -0.299 & 0.040 & 0.485 & -0.469 & -0.011 & 1.365 & 0.329 & 1.485 \\
NMM-LEVY & -0.006 & 0.014 & -0.020 & 0.067 & -0.296 & 0.041 & 0.476 & -0.487 & -0.012 & 1.027 & 0.372 & 1.312 \\
NMM-LEVY-E & -0.012 & 0.012 & -0.024 & 0.065 & -0.374 & 0.041 & 0.549 & -0.595 & -0.013 & 0.906 & 0.359 & 1.267  \\
\hline
\end{tabular}
\begin{tablenotes}[para,flushleft]
\small
\item[a] Best performance is in bold. 
\item[b] No comparison for volatility and downside deviation since every portfolio is scaled to the same target annualised volatility in \ref{definition:position_signal} for direct comparison of the net Sharpe.
\end{tablenotes}
\end{threeparttable}}
\end{table}

Based on the data in Panel A of Table \ref{tab:short_performance}, the benchmark model MACD averages a loss in short positions on the bootstrapped dataset, with a net Sharpe of $-0.396$ and the highest MDD across both bootstrapped and real price data. In contrast, NMM models improve performance in short positions by reducing losses. Specifically, NMM-DDTW-E enhances performance over MACD by reducing losses by 35\% and increasing the net Sharpe ratio by 24\% on bootstrapped data. It also achieves the highest Sortino and Calmar ratios, indicating effective downside risk and MDD control. Despite MACD's higher hit rate in short positions, its skewness score of 0.804 is lower than that of NMM-DDTW-E, which scores 1.155, and other NMM models. This indicates that NMM models not only result in smaller losses but also achieve more substantial occasional gains.

In Panel B of Table \ref{tab:short_performance}, NMM models continue to demonstrate effective loss control in short positions on the real price data. NMM-SDTW-E and NMM-DTW-E notably improve net Sharpe and reduce MDD to the greatest extent compared to the benchmark, respectively, with NMM-DDTW-E again achieving the most positively skewed performance, mirroring its success on bootstrapped datasets.

In the long direction, as detailed in Table \ref{tab:long_performance}, MACD demonstrates strong profitability with a net Sharpe ratio of 0.559 with the highest hit rate at 0.554. Among the network momentum models, NMM-LEVY outperforms with a net Sharpe of 0.587, a 6.1\% increase over the benchmark. It also reduces the MDD to 0.168, indicating superior loss control. Notably, while some network momentum models exhibit slightly lower net Sharpe ratios in long positions compared to the benchmark, all of them demonstrate more positively skewed returns, signifying smaller average losses and occasional larger gains. NMM-SDDTW achieves the most positively skewed returns, with a 76.6\% increase over MACD's skewness. This highlights the robust capability of network momentum models in long positions. Corresponding performance on the real price data in Panel B of Table \ref{tab:long_performance} further supports this, showing that NMM-LEVY has a higher Sharpe and Sortino ratio compared to the benchmark, and NMM-SDDTW-E records the most skewed returns.

\begin{table}[htbp]
\centering
\caption{Performance Metrics for Various Signals in Long Direction Only}
\label{tab:long_performance}
\resizebox{\textwidth}{!}{%
\begin{threeparttable}
\begin{tabular}{lcccccccccccc}
\hline
& Gross & Transaction & Net & vol. & Sharpe & downside & MDD & Sortino & Calmar  & Skewness & hit rate & Avg. P  \\
& Return & & Return & & &deviation & & & & & & Avg. L  \\
\hline
\multicolumn{11}{c}{\textbf{Panel A: Average Performance on 100 Bootstrapped Price Data}} \\
\hline
MACD & 0.068 & 0.012 & 0.055 & 0.099 & 0.559 & 0.057 & 0.191 & 0.983 & 0.091 & 0.367 & \textbf{0.554} & 1.243 \\
NMM-DTW & 0.069 & 0.015 & 0.054 & 0.100 & 0.540 & 0.054 & 0.186 & 0.998 & 0.093 & 0.553 & 0.519 & 1.412 \\
NMM-DTW-E & 0.069 & 0.012 & 0.057 & 0.099 & 0.572 & 0.054 & 0.172 & 1.053 & 0.103 & 0.574 & 0.519 & 1.451  \\
NMM-DDTW & 0.070 & 0.015 & 0.055 & 0.101 & 0.542 & 0.055 & 0.192 & 1.001 & 0.090 & 0.565 & 0.522 & 1.401  \\
NMM-DDTW-E & 0.068 & 0.012 & 0.056 & 0.099 & 0.568 & 0.054 & 0.173 & 1.053 & 0.101 & 0.594 & 0.518 & 1.459  \\
NMM-SDTW & 0.071 & 0.014 & 0.057 & 0.101 & 0.563 & 0.055 & 0.188 & 1.047 & 0.098 & 0.557 & 0.525 & 1.405  \\
NMM-SDTW-E & 0.068 & \textbf{0.011} & 0.056 & 0.099 & 0.569 & 0.054 & 0.171 & 1.056 & 0.102 & 0.602 & 0.518 & 1.460  \\
NMM-SDDTW & 0.069 & 0.015 & 0.053 & 0.100 & 0.529 & 0.054 & 0.193 & 0.997 & 0.088 & \textbf{0.648} & 0.517 & 1.417  \\
NMM-SDDTW-E & 0.068 & 0.012 & 0.056 & 0.099 & 0.569 & 0.053 & 0.172 & 1.066 & 0.102 & 0.648 & 0.517 & \textbf{1.468}  \\
NMM-LEVY & \textbf{0.072} & 0.013 & \textbf{0.059} & 0.100 & \textbf{0.587} & 0.055 & \textbf{0.168} & \textbf{1.076} & 0.109 & 0.477 & 0.534 & 1.377  \\
NMM-LEVY-E & 0.069 & 0.012 & 0.057 & 0.098 & 0.582 & 0.054 & 0.161 & 1.076 & \textbf{0.110} & 0.542 & 0.527 & 1.415  \\
\hline
\multicolumn{11}{c}{\textbf{Panel B: Performance on Real Price Data}} \\
\hline 
MACD & 0.064 & 0.012 & 0.052 & 0.094 & 0.557 & 0.049 & 0.199 & 1.065 & 0.076 & 0.623 & \textbf{0.547} & 1.276  \\
NMM-DTW & 0.062 & 0.014 & 0.047 & 0.096 & 0.494 & 0.048 & 0.216 & 0.995 & 0.063 & 0.901 & 0.500 & 1.479 \\
NMM-DTW-E & 0.067 & 0.011 & 0.055 & 0.095 & 0.581 & 0.047 & 0.158 & 1.170 & 0.100 & 0.918 & 0.491 & \textbf{1.649}  \\
NMM-DDTW & 0.065 & 0.014 & 0.051 & 0.098 & 0.519 & 0.050 & 0.214 & 1.011 & 0.069 & 0.831 & 0.496 & 1.531  \\
NMM-DDTW-E & 0.062 & 0.011 & 0.051 & 0.095 & 0.535 & 0.048 & 0.193 & 1.052 & 0.076 & 0.868 & 0.517 & 1.418 \\
NMM-SDTW & 0.070 & 0.013 & 0.057 & 0.095 & 0.600 & 0.046 & \textbf{0.153} & 1.236 & \textbf{0.108} & 0.916 & 0.513 & 1.538  \\
NMM-SDTW-E & 0.062 & \textbf{0.011} & 0.051 & 0.094 & 0.543 & 0.047 & 0.170 & 1.088 & 0.087 & 0.904 & 0.500 & 1.545  \\
NMM-SDDTW & 0.061 & 0.014 & 0.046 & 0.098 & 0.470 & 0.047 & 0.220 & 0.972 & 0.060 & 0.850 & 0.500 & 1.441  \\
NMM-SDDTW-E & 0.065 & 0.011 & 0.053 & 0.095 & 0.565 & 0.047 & 0.168 & 1.139 & 0.092 & \textbf{0.935} & 0.504 & 1.550  \\
NMM-LEVY & \textbf{0.070} & 0.012 & \textbf{0.058} & 0.094 & \textbf{0.610} & 0.046 & 0.169 & \textbf{1.257} & 0.098 & 0.810 & 0.526 & 1.449\\
NMM-LEVY-E & 0.068 & 0.011 & 0.056 & 0.094 & 0.597 & 0.047 & 0.155 & 1.187 & 0.105 & 0.813 & 0.517 & 1.495  \\
\hline
\end{tabular}
\begin{tablenotes}[para,flushleft]
\small
\item[a] Best performance is in bold. 
\item[b] No comparison for volatility and downside deviation since every portfolio is scaled to the same target annualised volatility in \ref{definition:position_signal} for direct comparison of the net Sharpe.
\end{tablenotes}
\end{threeparttable}}
\end{table}

\subsection{Diversification Analysis}
\begin{figure}[htbp]
    \centering
    \begin{subfigure}{.5\textwidth}
        \centering
        \includegraphics[scale=.5]{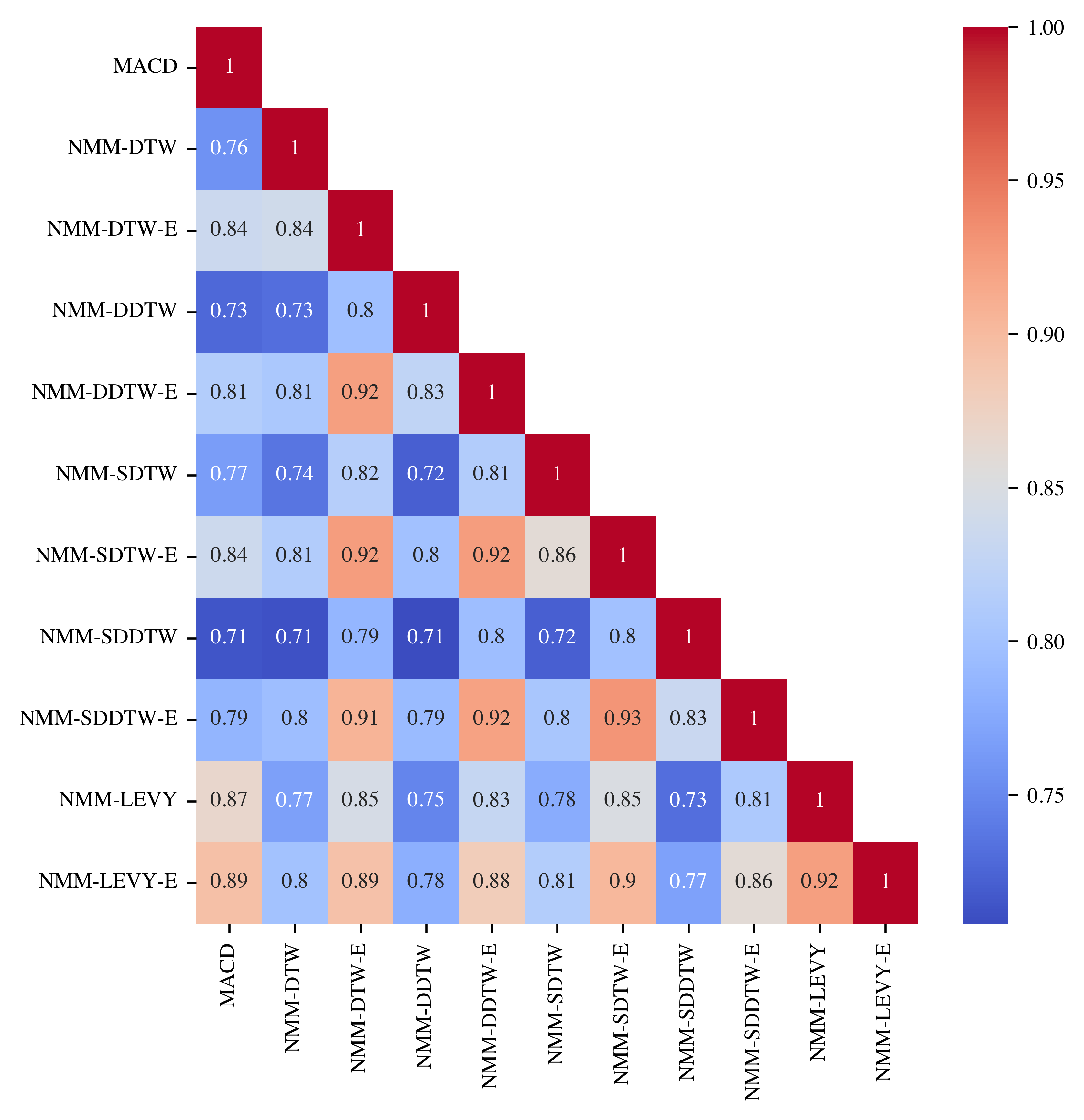}
        \caption{}
        \label{fig:pnl_corr_a}
    \end{subfigure}%
    \begin{subfigure}{.5\textwidth}
        \centering
        \includegraphics[scale=.5]{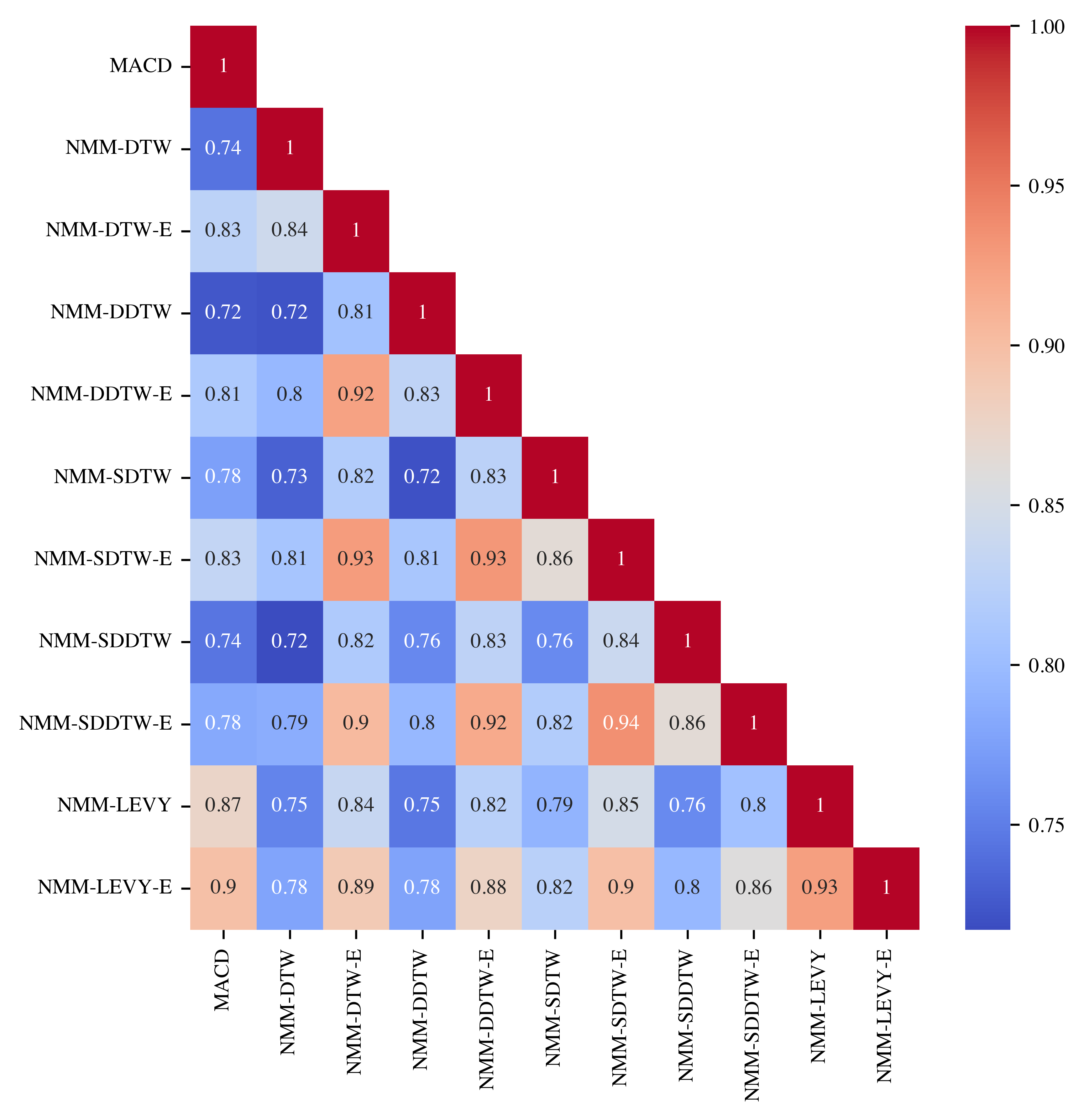}
        \caption{}
        \label{fig:pnl_corr_b}
    \end{subfigure}
    \caption{A diversification analysis on the PnL pairwise correlation between models on the bootstrapped datasets (left) and real price dataset (right).}
    \label{fig:pnl_corr}
\end{figure}

We analyse the correlation of their returns to assess whether the NMM models and MACD exhibit orthogonal trading signals. Figure \ref{fig:pnl_corr_a} presents the average correlation on bootstrapped datasets, while Figure \ref{fig:pnl_corr_b} the correlation on real price data covering the entire out-of-sample period from 2005 to 2024.

By analysing the returns between the NMM models and the benchmark MACD on the bootstrapped datasets, we notice that the average correlations range from 0.71 to 0.89 in Figure \ref{fig:pnl_corr_a}. NMM-SDDTW exhibits the lowest average correlation with MACD at 0.71, similar to NMM-DDTW's correlation with MACD. NMM-LEVY and NMM-LEVY-E show slightly higher correlations with MACD at 0.87 and 0.89, respectively. Comparable results are observed in the PnL from the real price data between 2005 and 2024 in Figure \ref{fig:pnl_corr_b}, where NMM-DDTW and NMM-SDDTW display the lowest correlations with MACD, at 0.72 and 0.74, respectively. Although the PnL correlations are not completely orthogonal, these empirical findings support the existence of additional information captured in our NMM models.

Our empirical findings indicate that different DTW algorithms capture distinct lead-lag relationships, consequently influencing the network momentum identified. Specifically, NMM models employing multi-dimensional DTW algorithms, such as NMM-SDTW and NMM-SDDTW, exhibit lower correlation values, around 0.7, with models based on one-dimensional DTW algorithms like NMM-DTW and NMM-DDTW in Figure \ref{fig:pnl_corr_a}. This suggests that multi-dimensional DTW effectively captures different lead-lag relationships with one-dimensional approaches. Furthermore, NMM-LEVY demonstrates correlations ranging from 0.73 to 0.83 with NMM-DTW and its variations, indicating that using the L\'evy area as a lead-lag detection method yields additional results from those obtained via dynamic time warping algorithms.

It is also noteworthy that correlations between each NMM model and its ensemble variant range from 0.80 to 0.92. This implies that while there is some dependency, the ensemble method still introduces different information on the lead-lag relationship. This is achieved by utilising six different lookback lengths, leading to variations in the network momentum model outcomes. The ensemble approach thus contributes uniquely to understanding and leveraging network momentum in trading strategies.

\begin{figure}[htbp]
    \centering
    \begin{subfigure}{.5\textwidth}
        \centering
        \includegraphics[scale=0.5]{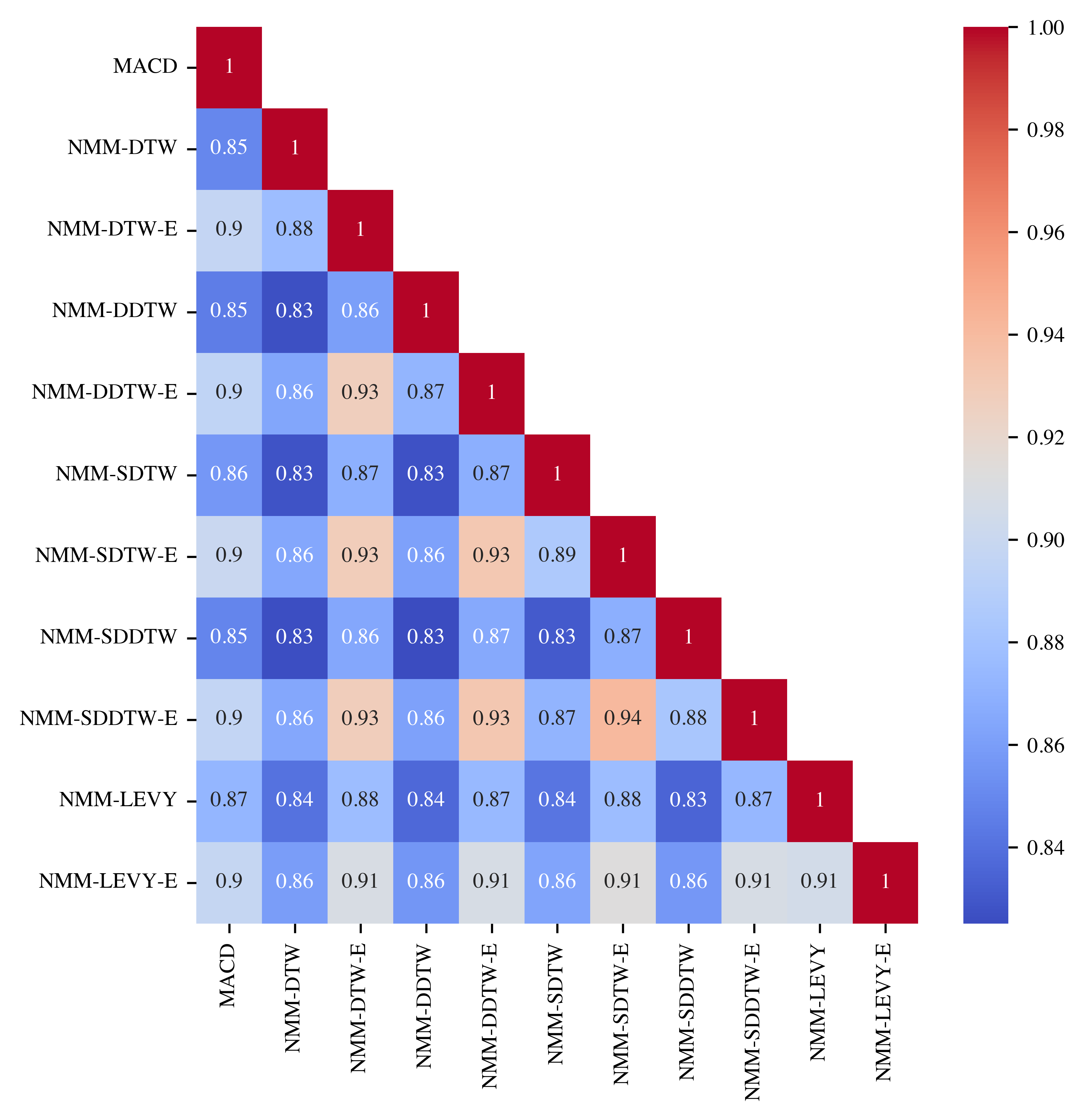}
        \caption{} 
        \label{fig:sign_agreement_a}
    \end{subfigure}%
    \begin{subfigure}{.5\textwidth}
        \centering
        \includegraphics[scale=0.5]{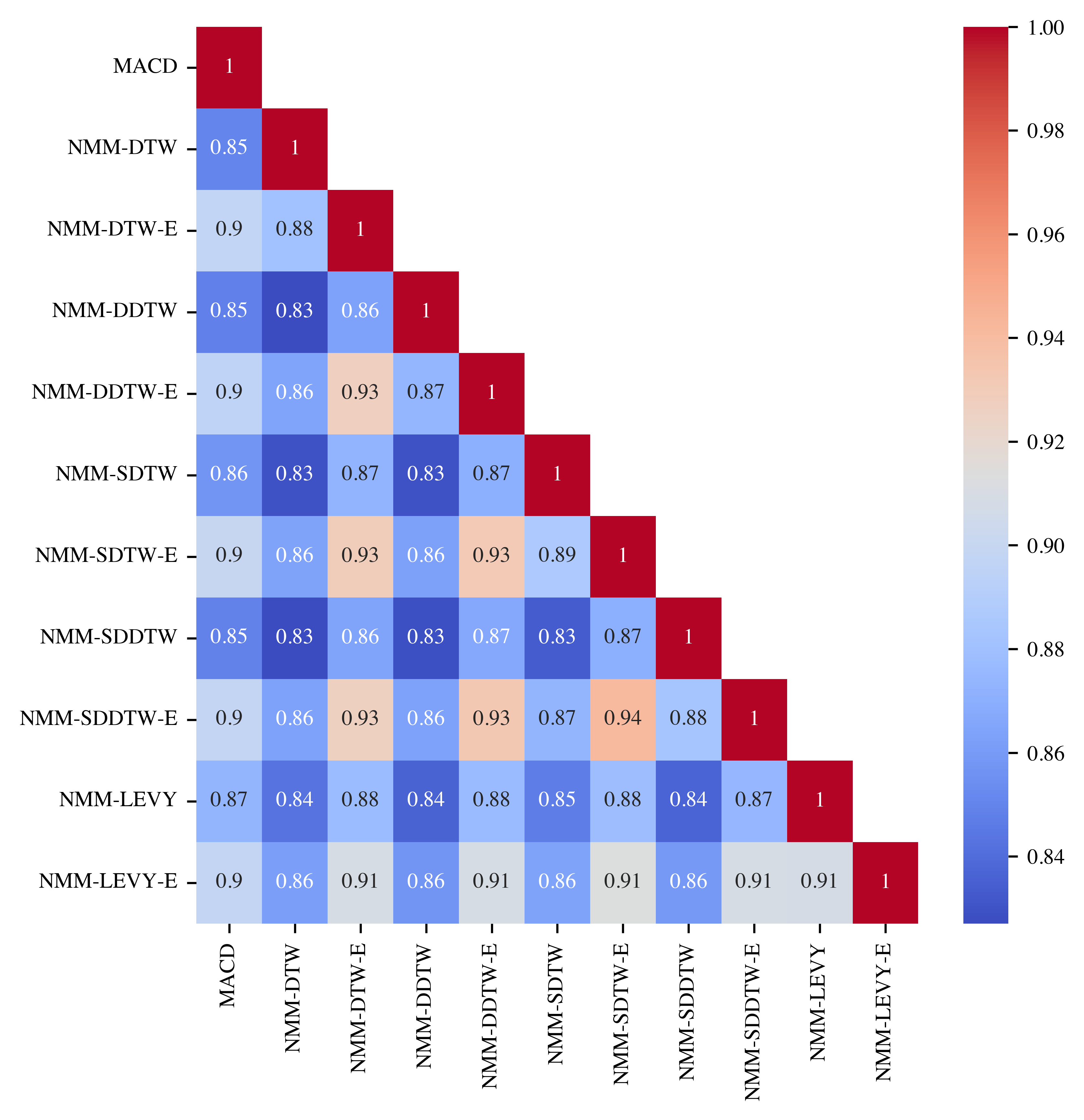}
        \caption{} 
        \label{fig:sign_agreement_b}
    \end{subfigure}
    \caption{A diversification analysis on the pairwise sign agreement between models on the bootstrapped datasets (left) and real price dataset (right)}
    \label{fig:sign_agreement}
\end{figure}

Next, we introduce a second metric for our diversification analysis: the sign agreement between two models. This metric is the percentage of days on which two models share the same trading direction—either opting to go long or short on the market on a trading day—across the entire portfolio. The average results on bootstrapped data is presented in Figure \ref{fig:sign_agreement_a}, with performance on real price data from the entire out-of-sample period from 2005 to 2024 in Figure \ref{fig:sign_agreement_b}. We also examine the average annualised expected PnL on days when the NMM models diverge in sign from the benchmark MACD model to assess whether differences in trading direction result in additional profits. The differences in average profits between the models (network momentum models minus MACD) for these days are detailed in Table~\ref{tab:sign_diff_pnl}.

\begin{table}[htbp]
\centering
\caption{Average PnL Gains Over Benchmark on Opposing Signal Days}
\label{tab:sign_diff_pnl}
\resizebox{\textwidth}{!}{%
\begin{tabular}{lcccccccccc}
\hline
& NMM-DTW & NMM-DTW-E & NMM-DDTW & NMM-DDTW-E & NMM-SDTW & NMM-SDTW-E & NMM-SDDTW & NMM-SDDTW-E & NMM-LEVY & NMM-LEVY-E \\
\hline
Bootstrapped data & 0.011 & \textbf{0.032} & 0.001 & 0.023 & -0.002 & 0.021 & -0.020 & 0.026 & 0.013 & 0.004 \\
Real Price data & 0.017 & 0.043 & -0.017 & 0.015 & \textbf{0.057} & -0.008 & -0.063 & 0.039 & 0.051 & 0.031 \\
\hline
\end{tabular}}
\end{table}

It can be observed that NMM-DTW and NMM-DDTW, with the lowest sign agreement with the MACD at 85\%, result in average additional returns of 0.011 and 0.032, respectively. Most other NMM models show a sign agreement ranging from 85\% to 90\% with MACD and generally yield higher returns than MACD on days with differing signs, except for NMM-SDTW and NMM-SDDTW, which achieve less profits than the benchmark on these days. These empirical results suggest that the additional network momentum captured by the NMM models is effective at following and adjusting to trends identified by the MACD, which focuses solely on time-series momentum. This indicates that our models are robust and effective in identifying network momentum within a portfolio.

On the real price data from 2005 to 2024, it is notable that the NMM-SDTW achieves the highest average returns gain over the benchmark model with an annualised expected difference at 0.057,  with a sign agreement of 86\%. However, NMM-DDTW, NMM-SDTW-E, and NMM-SDDTW realise lower profits compared to MACD, with respective losses of -0.017, -0.008 and -0.063, respectively, , and sign agreements of 85\%, 90\%, and 85\%.

\subsection{Skewness Analysis}\label{skewness_analysis}
In this final section on performance analysis, we examine the skewness of returns from NMM models across different time horizons and compare them with the benchmark MACD model. 

As highlighted in \cite{martin2023design}, effective trend-following strategies often exhibit a long-option-type payoff, attributed to positive skewness. This phenomenon can be conceptualised as the purchase of an option: regular small losses represent the premium paid, while correctly identifying and riding a trend may result in significant gains, analogous to an option's payoff.

\begin{figure}[htbp]
    \centering
    \scalebox{1.0}{
    \begin{minipage}{0.49\textwidth}
        \centering
        \includegraphics[width=\textwidth]{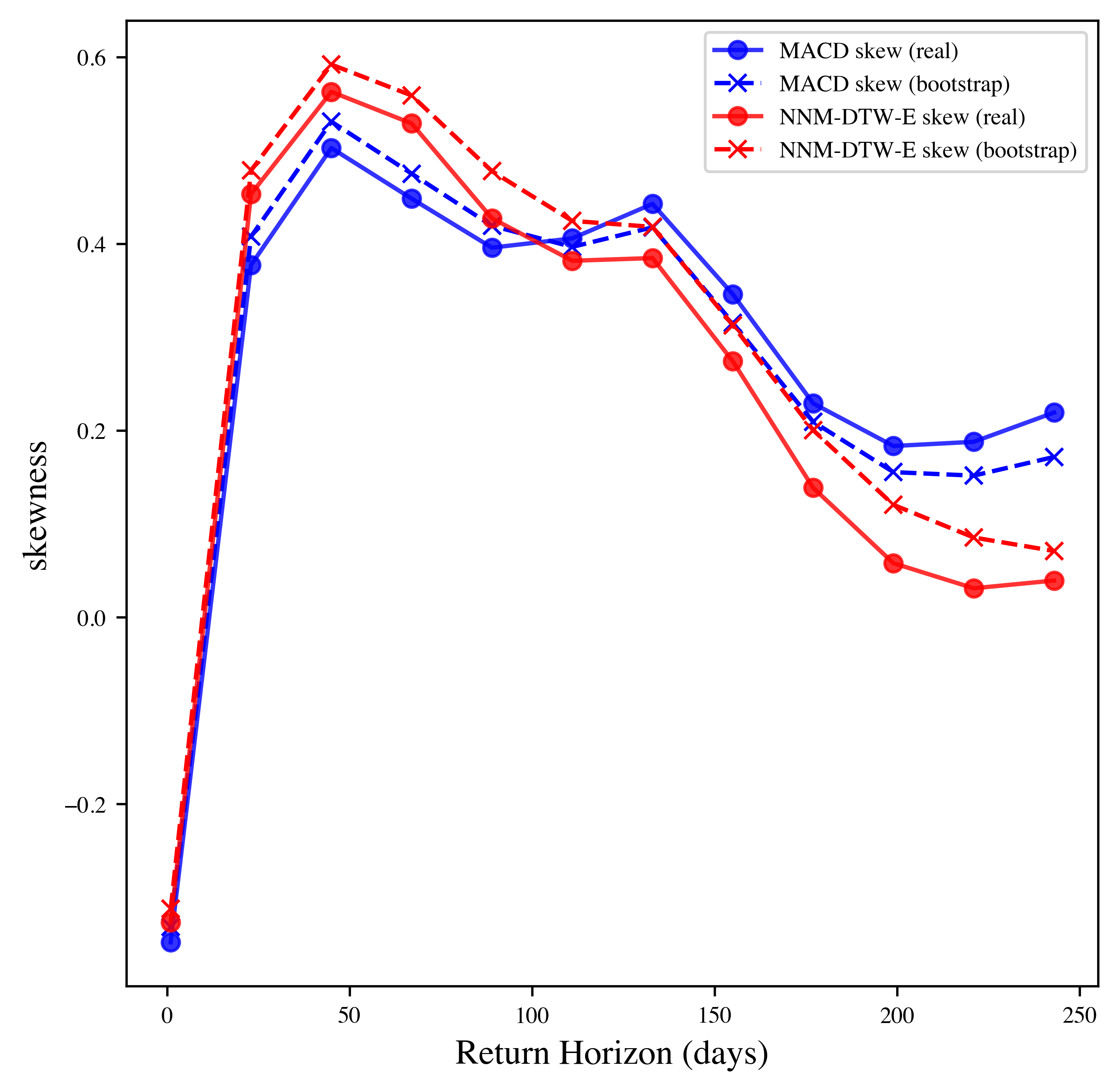}
        \caption*{(a)}
    \end{minipage}\hfill
    \begin{minipage}{0.49\textwidth}
        \centering
        \includegraphics[width=\textwidth]{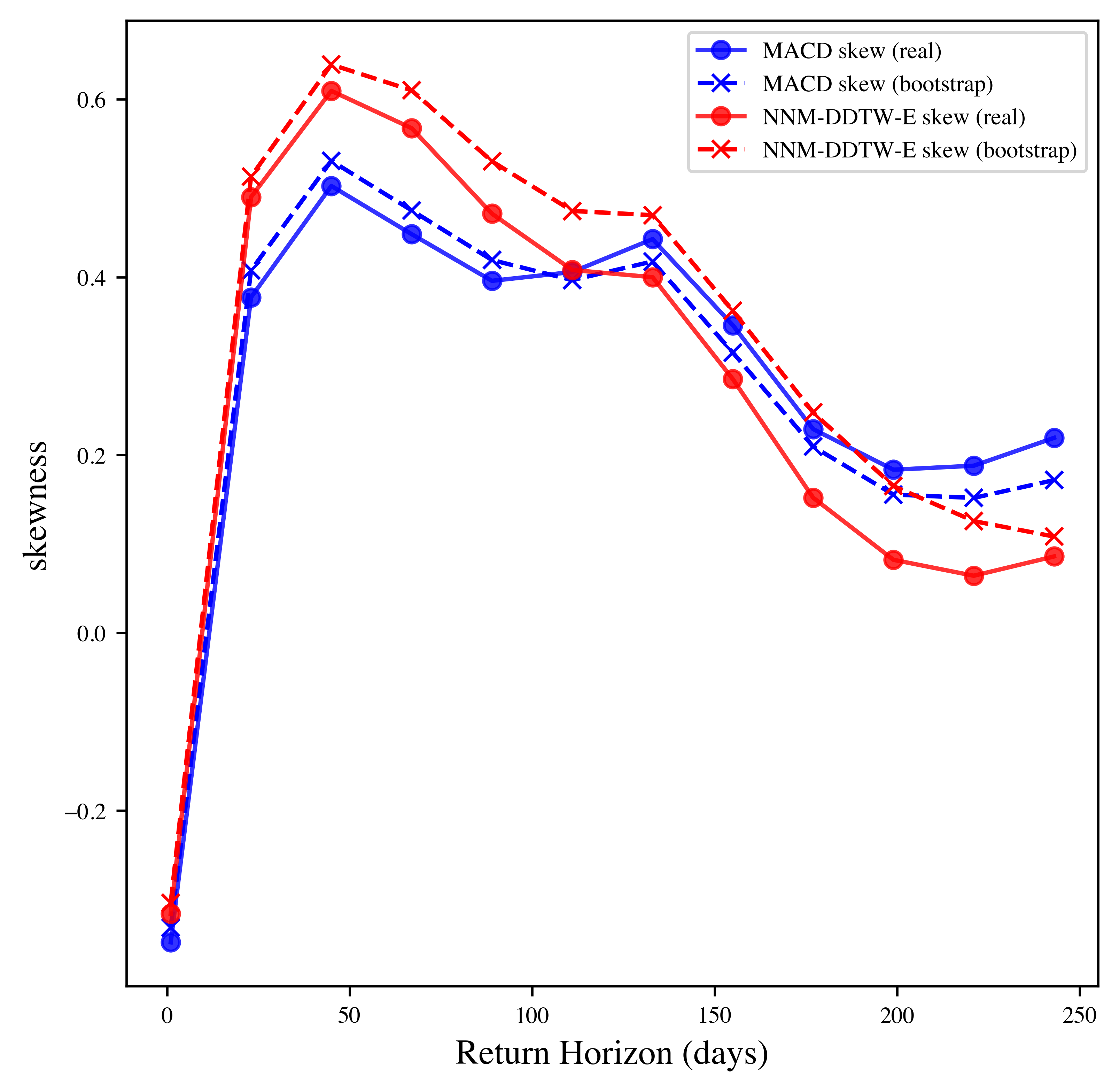}
        \caption*{(b)}
    \end{minipage}
    }
    \vspace{0.05cm} 
    \scalebox{1.0}{
    \begin{minipage}{0.49\textwidth}
        \centering
        \includegraphics[width=\textwidth]{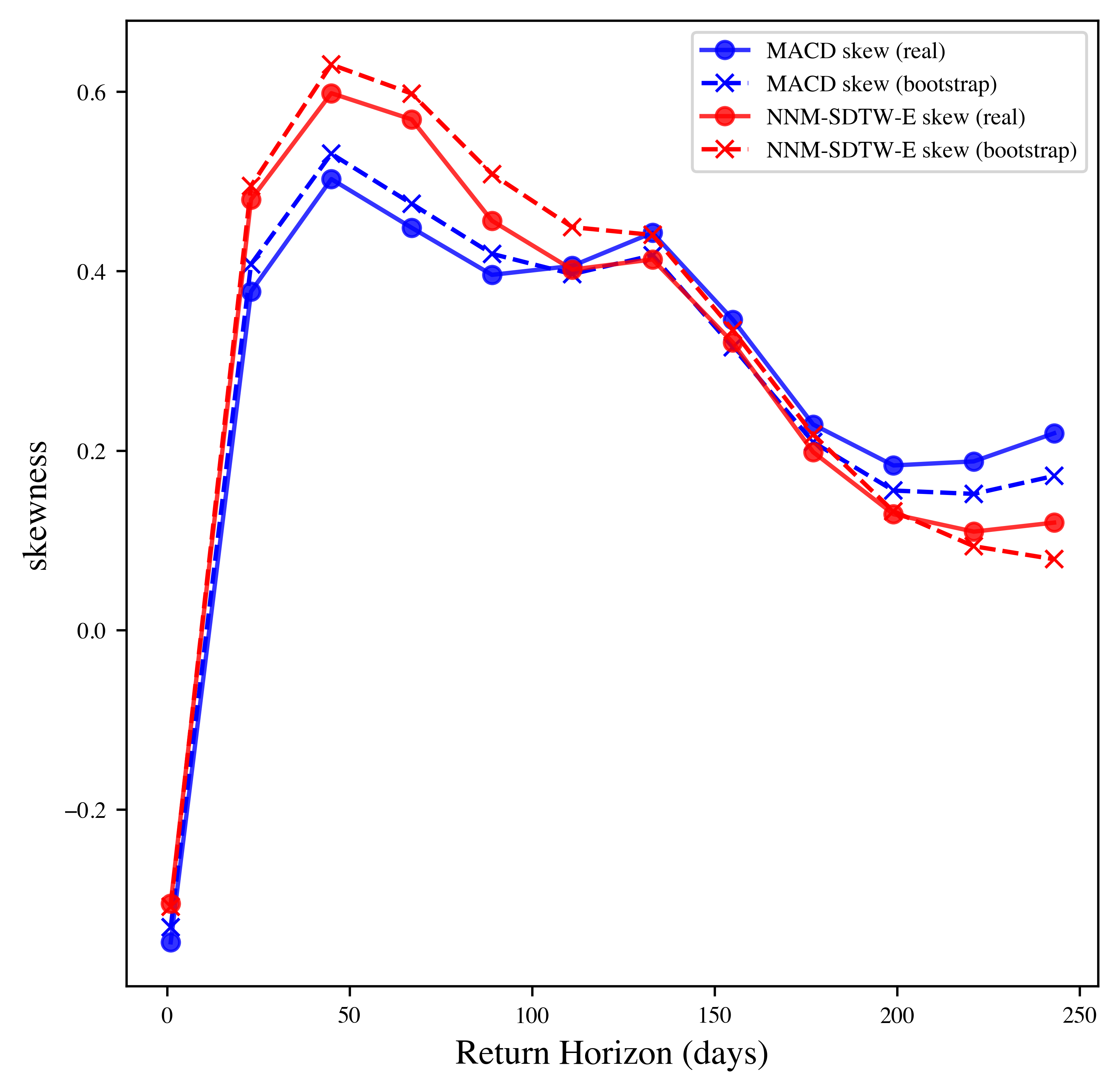}
        \caption*{(c)}
    \end{minipage}\hfill
    \begin{minipage}{0.49\textwidth}
        \centering
        \includegraphics[width=\textwidth]{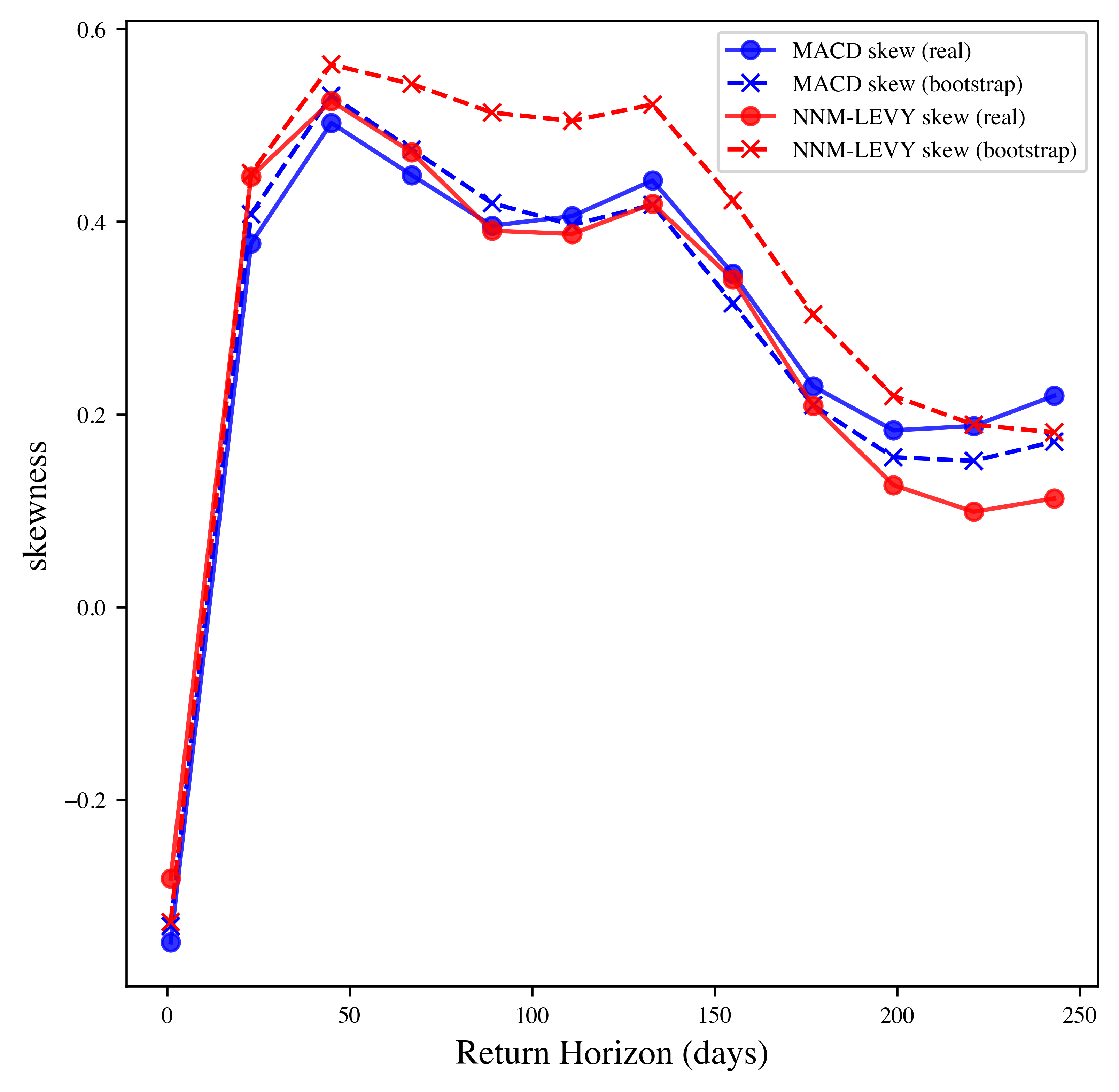}
        \caption*{(d)}
    \end{minipage}
    }
    \caption{Skewness in the returns of the network momentum model over various periods, compared to those of the time series momentum model, using different lead-lag detection models: (a) NMM-DTW-E, (b) NMM-DDTW-E, (c) NMM-SDTW-E, and (d) NMM-LEVY.}
    \label{fig:network_model_skewness}
\end{figure}

We present the skewness across various return horizons for four NMM models in Figure~\ref{fig:network_model_skewness} for detailed analysis. The four representative network momentum models are: NMM-DTW-E (a), which performs the best on real price data and achieves the highest average PnL gain over the benchmark on days with opposing signals; NMM-DDTW-E (b), the top performer on bootstrapped data and in short positions; NMM-SDTW-E (c), which outperforms the other multi-dimensional dynamic time warping models; and NMM-LEVY (d), the top performer in long positions. Our empirical study finds that the skewness of the network momentum models exhibits a similar and consistent pattern; therefore, we only include four examples here. For completeness, we include plots for the other NMM models in Appendix~\ref{app:supplimentary_plot}.

Our analysis indicates that the NMM models exhibit stronger positive skewness in returns across time horizons ranging from days to months compared to the benchmark MACD, with a notable peak at the two-month return horizon. This suggests that the NMM models are more effective at identifying and positioning for trends ahead of time to capitalise on these opportunities. Even in the case of daily returns, where all models typically exhibit a negative skew due to the option-like payoff characteristic of trend-following strategies, the NMM models show less negative skewness, indicating better risk control and the ability to identify short-term trends without enduring prolonged periods of losses.

However, it is important to note that the NMM models demonstrate a more pronounced decay in skewness over longer horizons compared to MACD. Particularly from half-year to one-year return horizons, although the NMM models still maintain positive skewness, it is less pronounced than that observed with MACD.

The pattern of skewness across different time horizons aligns with the findings reported in \cite{martin2023design}, which we refer interested readers to for further details. Our empirical results suggest that NMM models not only uphold the desired characteristics of a trend-following strategy but also enhance them. They effectively capture network momentum spillover and identify both short-term and medium-term trends accurately, thereby enabling the models to anticipate market movements by considering momentum from interconnected markets within the portfolio.

\section{Conclusion}\label{conclusion}
We propose a methodology that transforms cross-sectional momentum spillover into network momentum across market industries. This process utilises two lead-lag detection models to identify non-linear relationships at fixed lags and between non-synchronised market returns. We then apply a graph learning model to quantify the intricate interconnectedness of market leadership and individual momentum, generating a novel trading signal. This signal is utilised to construct a portfolio for a systematic trend-following strategy, which we evaluate using 100 sets of bootstrapped price data from 28 futures contracts across metals, agriculture, energy, and equities. We backtest our strategy in a realistic trading environment that accounts for time delays in establishing positions.

Our framework enhances the performance of traditional trend-following strategies, consistently achieving a higher and statistically significant net Sharpe ratio compared to time series momentum strategies. Our model also robustly reduces transaction costs and enhances performance over time series momentum strategies in short positions, where the latter typically incurs losses. By employing various lead-lag detection techniques, our network momentum models generate low-correlated signals that more effectively identify market trends by establishing positions in the correct direction. The proposed framework also consistently yields more positively skewed returns, underscoring the efficiency and robustness of the network momentum identified for trend-following strategies. 

Most importantly, the results indicate that the superior performance of converting cross-sectional momentum into network momentum is not confined to specific market combinations within the portfolio, nor is it dependent on historical market trends. Instead, the proposed network momentum model demonstrates remarkable generalisability across various industries and markets.

We propose several future research directions. Firstly, exploring non-linear ensemble methods on the lead-lag matrices computed by multiple models could be beneficial. Considering that the divergence analysis indicates dynamic time warping and L\'evy area models capture different information, their combination in a non-linear manner may enhance the identification of lead-lag relationships. Secondly, investigating asymmetrical adjacency matrices with machine learning models like graph neural networks could shed light on potential non-symmetrical relationships between markets. Thirdly, while our current portfolio construction combines time series momentum features with equal weights and applies the same adjacency matrix to all of them, it may be worthwhile to explore fitting different lead-lag matrices and adjacency matrices to time series momentum features at varying speeds. Employing non-linear methods to combine these may more effectively capture the nonlinearity in momentum spillover.

\section*{Acknowledgments}

I would like to thank Dr William Ferreira, Leonardo Marroni, Irene Perdomo and Lorenzo Reati for the opportunity to undertake this project.

\section*{Appendix}
\subsection{Dataset Details}\label{app:used_markets}
In Table \ref{tab:used_markets}, we summarise the Bloomberg tickers and names of all the futures contracts we used in our portfolio.

\begin{table}[H]
\centering
\begin{tabular}{l l l}
\hline
\textbf{Bloomberg Ticker} & \textbf{Contract Name} & \textbf{Market Class} \\ \hline
future\_bo1\_comdty & CBOT Soybean Oil Future & Ags \\ \hline
future\_sm1\_comdty & CBOT Soybean Meal Future & Ags \\ \hline
future\_sb1\_comdty & NYBOT CSC Number 11 World Sugar Future & Ags \\ \hline
future\_rr1\_comdty & Rough Rice Future & Ags \\ \hline
future\_o\_1\_comdty & Oats Future & Ags \\ \hline
future\_mw1\_comdty & MGE Red Wheat Future & Ags \\ \hline
future\_kw1\_comdty & KCBT Hard Red Winter Wheat Future & Ags \\ \hline
future\_kc1\_comdty & NYBOT CSC C Coffee Future & Ags \\ \hline
future\_jo1\_comdty & Orange Juice (RTH) Future & Ags \\ \hline
future\_w\_1\_comdty & CBOT Wheat Future & Ags \\ \hline
future\_c\_1\_comdty & CBOT Corn Future & Ags \\ \hline
future\_cc1\_comdty & NYBOT CSC Cocoa Future & Ags \\ \hline
future\_da1\_comdty & Class III Milk Future & Ags \\ \hline
future\_ct1\_comdty & NYBOT CTN Number 2 Cotton Future & Ags \\ \hline
future\_cl1\_comdty & NYMEX Light Sweet Crude Oil Future & Energy \\ \hline
future\_co1\_comdty & ICE Brent Crude Oil Future & Energy \\ \hline
future\_ng1\_comdty & NYMEX Henry Hub Natural Gas Future & Energy \\ \hline
future\_cf1\_index & Euronext CAC 40 Index Future & Equity \\ \hline
future\_nq1\_index & CME E-Mini NASDAQ 100 Index Future & Equity \\ \hline
future\_vg1\_index & Eurex EURO STOXX 50 Future & Equity \\ \hline
future\_hi1\_index & HKG Hang Seng Index Future & Equity \\ \hline
future\_gx1\_index & Eurex DAX Index Future & Equity \\ \hline
future\_es1\_index & CME E-Mini Standard \& Poor's 500 Future & Equity \\ \hline
future\_pa1\_comdty & NYMEX Palladium Future & Metals \\ \hline
future\_pl1\_comdty & NYMEX Platinum Future & Metals \\ \hline
future\_hg1\_comdty & COMEX Copper Future & Metals \\ \hline
future\_si1\_comdty & COMEX Silver Future & Metals \\ \hline
future\_gc1\_comdty & COMEX Gold 100 Troy Ounces Future & Metals \\ \hline
\end{tabular}
\caption{Futures Contracts from Bloomberg}
\label{tab:used_markets}
\end{table}

\subsection{Supplementary Skewness Plots Across Time Horizons}\label{app:supplimentary_plot}
\begin{figure}[htbp]
    \centering
    \scalebox{1}{
    \begin{minipage}{0.5\textwidth}
        \centering
        \includegraphics[width=0.8\textwidth]{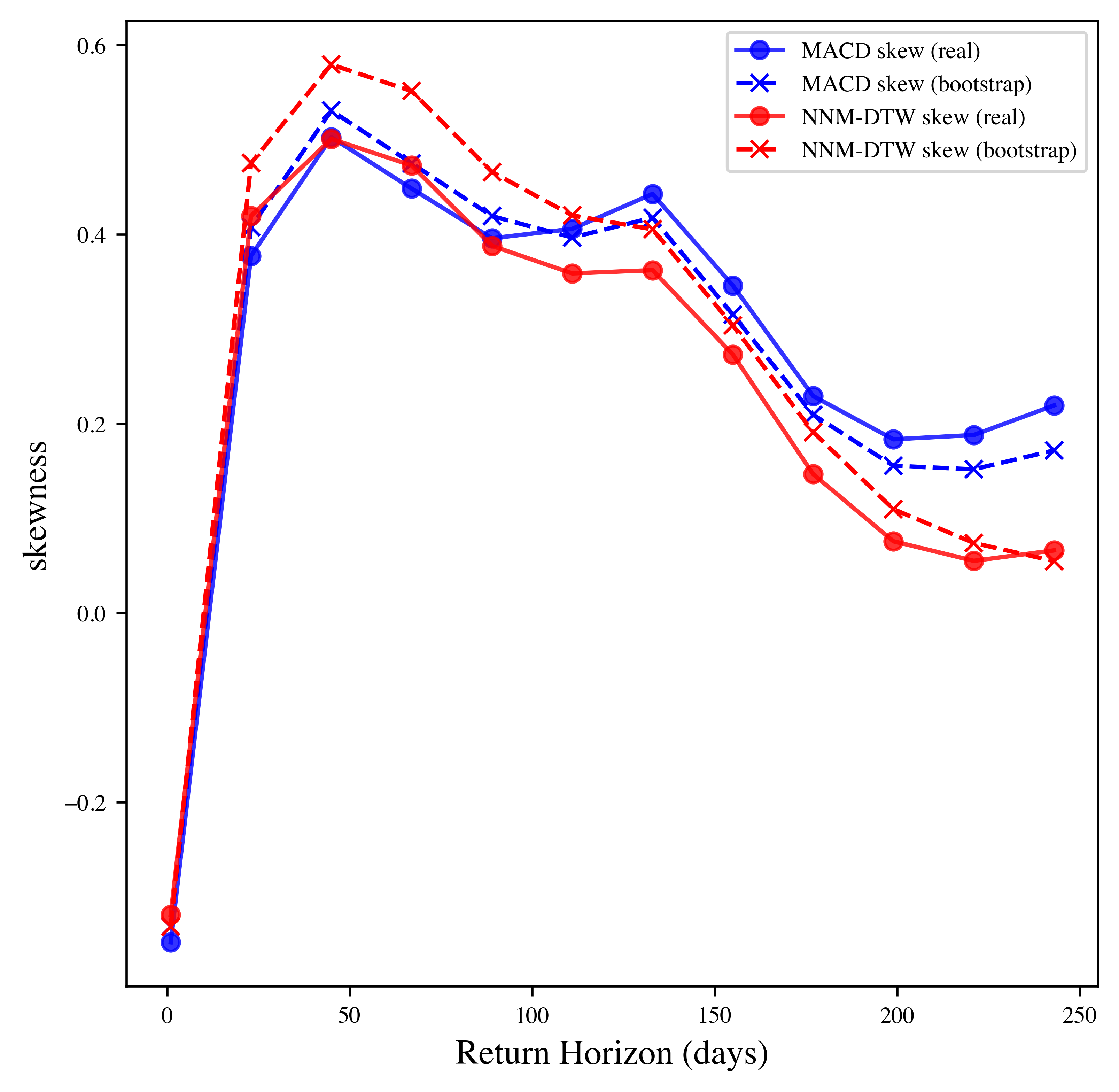} 
        \caption*{(a)}
    \end{minipage}\hfill
    \begin{minipage}{0.5\textwidth}
        \centering
        \includegraphics[width=0.8\textwidth]{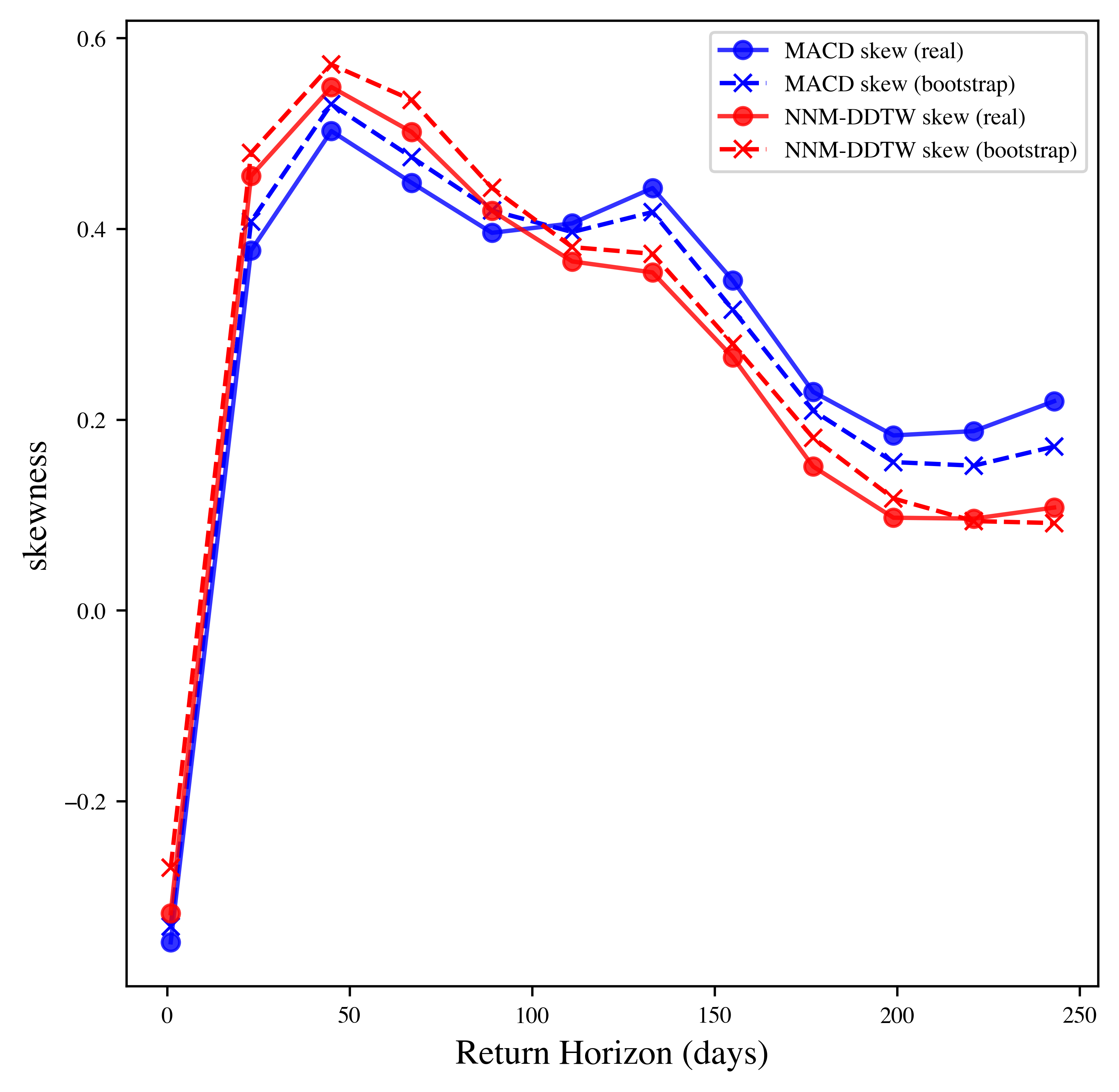}
        \caption*{(b)}
    \end{minipage}
    }
    \scalebox{1}{
    \begin{minipage}{0.5\textwidth}
        \centering
        \includegraphics[width=0.8\textwidth]{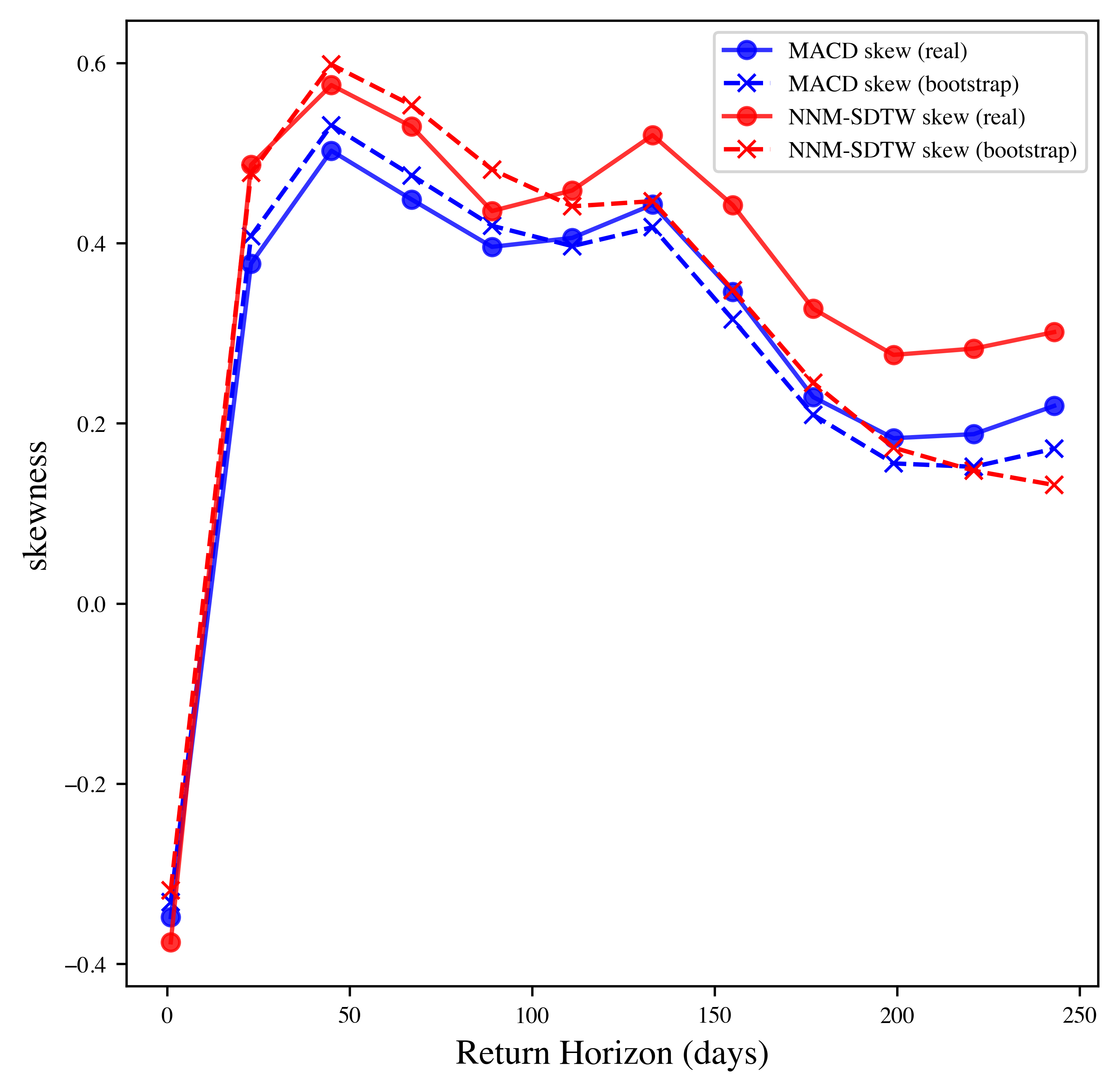}
        \caption*{(c)}
    \end{minipage}\hfill
    \begin{minipage}{0.5\textwidth}
        \centering
        \includegraphics[width=0.8\textwidth]{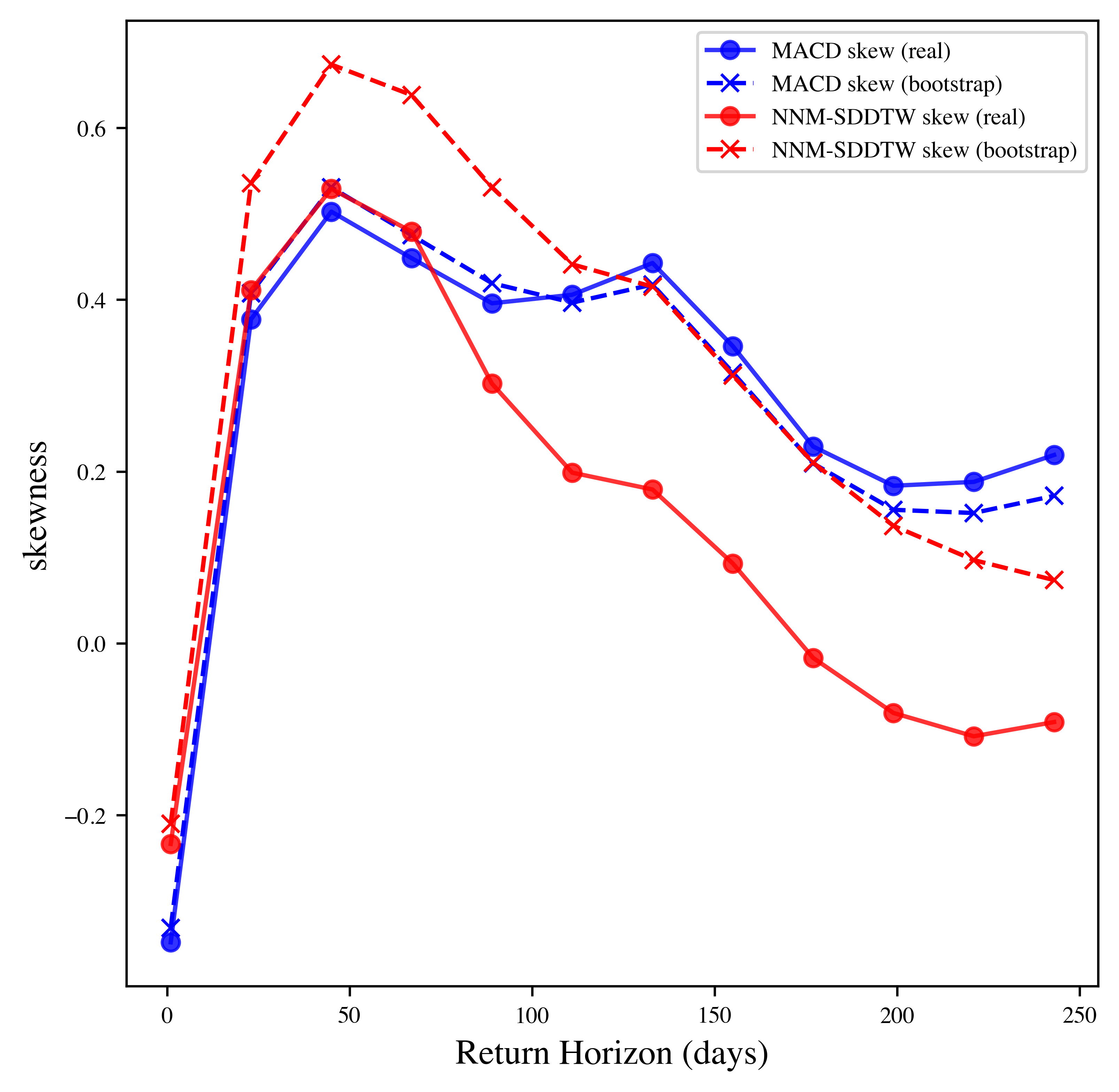}
        \caption*{(d)}
    \end{minipage}
    }
    \scalebox{1}{
    \begin{minipage}{0.5\textwidth}
        \centering
        \includegraphics[width=0.8\textwidth]{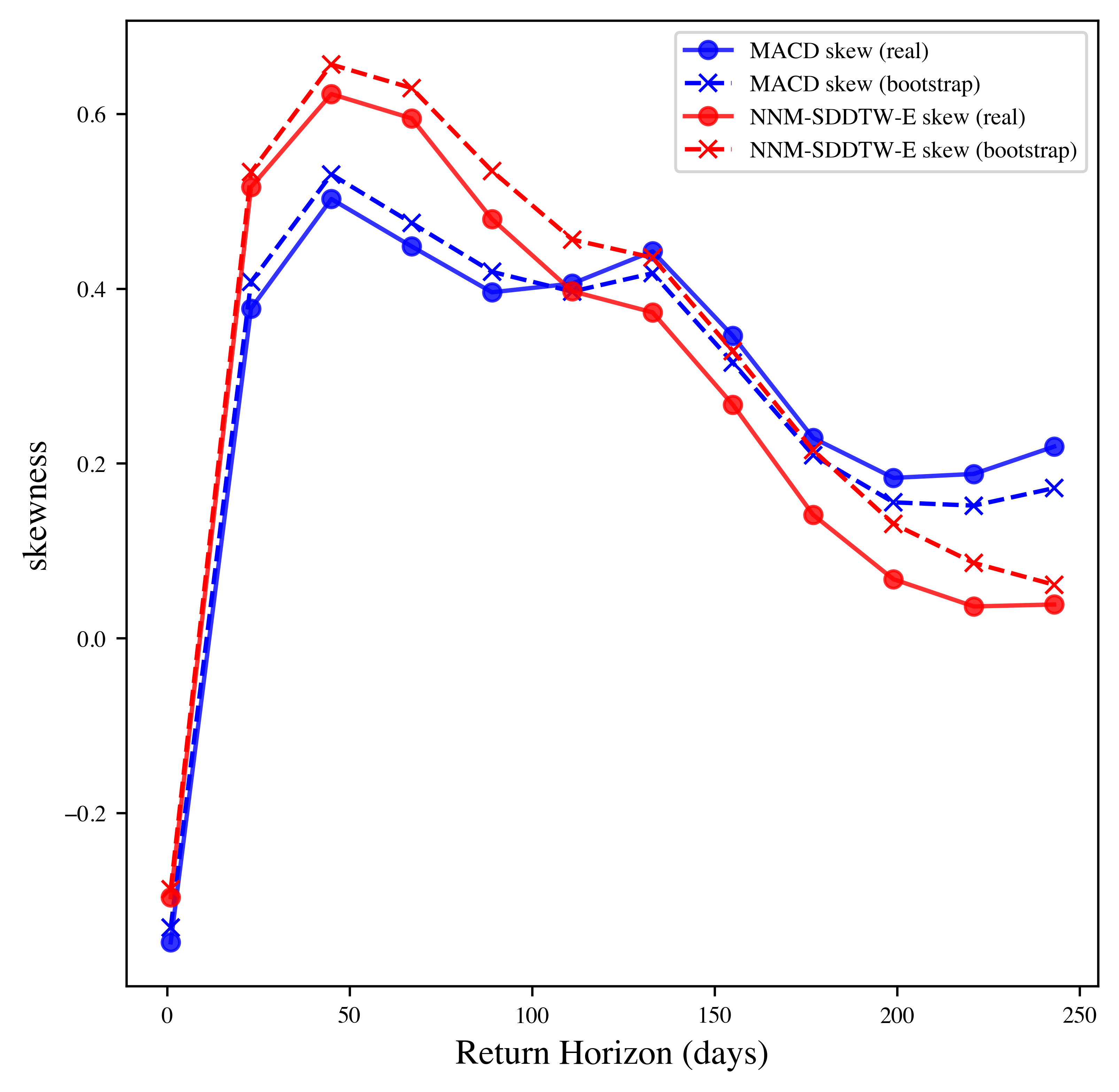}
        \caption*{(e)}
    \end{minipage}\hfill
    \begin{minipage}{0.5\textwidth}
        \centering
        \includegraphics[width=0.8\textwidth]{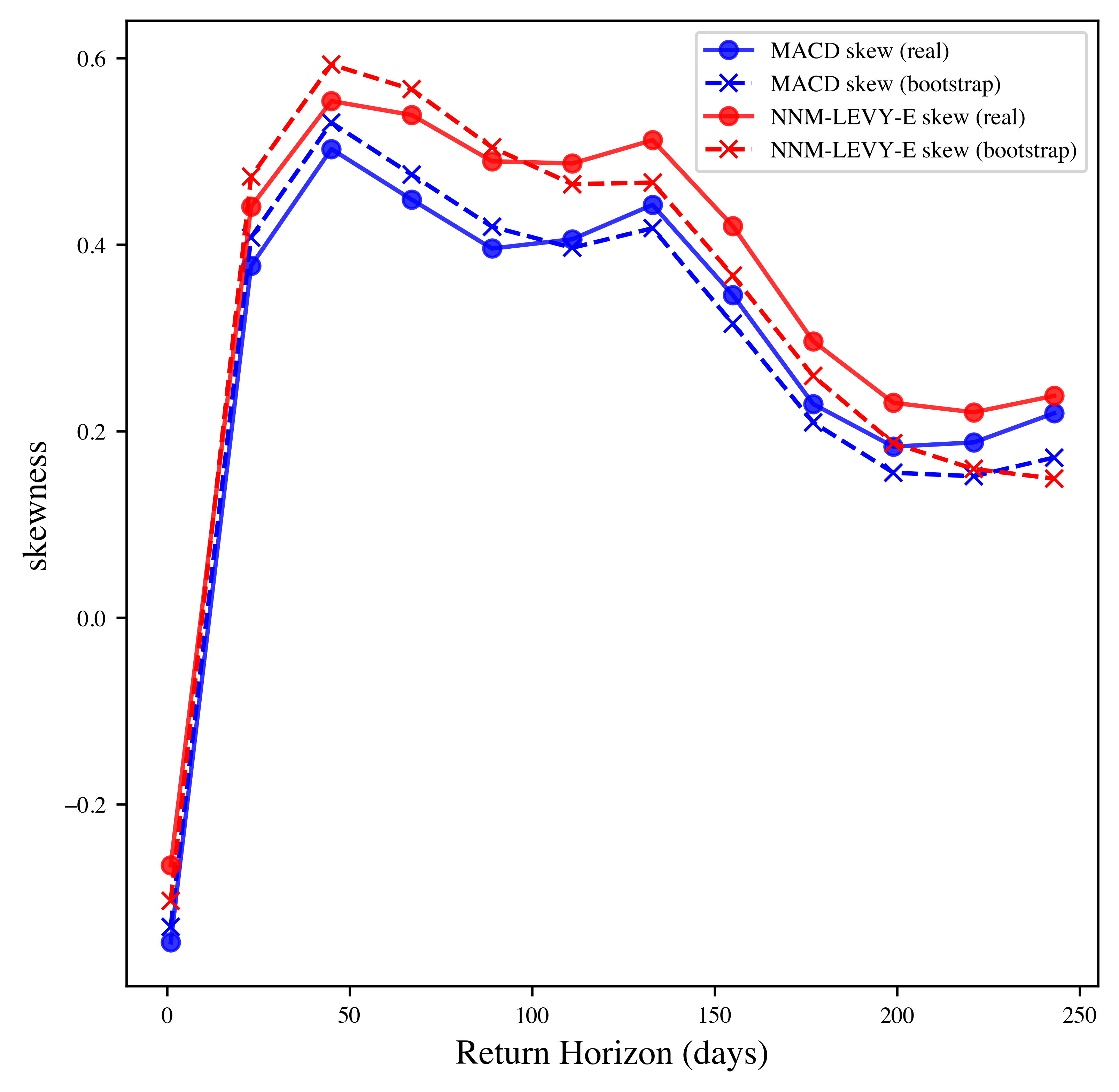}
        \caption*{(f)}
    \end{minipage}
    }
    \caption{Supplementary plots of skewness in the returns of the network momentum model over various periods, compared to those of the time series momentum model, using different lead-lag detection models.}
    \label{fig:network_model_skewness_supplementary}
\end{figure}

In Figure~\ref{fig:network_model_skewness_supplementary}, we include skewness plots for additional network momentum models not presented in Section~\ref{skewness_analysis}.

\newpage
\bibliographystyle{unsrt}  
\bibliography{references}  
\end{document}